\documentclass[useAMS,usenatbib]{mn2e}

\usepackage{psfig, epsf, epsfig}

\title[Galaxy evolution regulated by dust]{Dust-regulated galaxy formation and evolution:
A new chemodynamical model with live dust particles}
\author[K. Bekki]
{Kenji Bekki${}^1$\thanks{E-mail:
bekki@cyllene.uwa.edu.au} \\
${}^1$ICRAR M468
The University of Western Australia
35 Stirling Hwy, Crawley
Western Australia 6009, Australia}

\begin{document}

\date{Accepted, Received 2005 February 20; in original form }

\pagerange{\pageref{firstpage}--\pageref{lastpage}} \pubyear{2005}

\maketitle

\label{firstpage}

\begin{abstract}

Interstellar dust plays decisive roles in the conversion of neutral to molecular
hydrogen (${\rm H_2}$), 
the thermodynamical evolution of interstellar medium (ISM),
and the modification of spectral energy distributions (SEDs) of galaxies.
These important roles of dust  have not been self-consistently 
included in previous numerical simulations of
galaxy formation and evolution.
We have therefore developed a new  model by which one can investigate
whether and how galaxy formation and evolution can be 
influenced  by dust-related physical processes such as
photo-electric heating, ${\rm H_2}$ formation on dust,
and stellar radiation pressure on dust
in detail. A novel point of the model is that
different dust species in a galaxy are represented by `live dust' particles
(i.e., not test particles).
Therefore, dust particles in a galaxy not only interact gravitationally  with 
all four components of the galaxy
(i.e., dark matter, stars, gas, and dust) but also
are grown and destroyed  through physical processes of ISM.
First we describe a way to include dust-related physical processes
in Nbody+hydrodynamical simulations of galaxy evolution in detail.
Then we show some preliminary results of dust-regulated
galaxy evolution.
The preliminary results suggest that
the evolution of dust distributions driven by radiation
pressure of stars is very important for the  evolution of star formation
rates, chemical abundances,
${\rm H_2}$ fractions,  and gas distributions in galaxies.
\end{abstract}


\section{Introduction}

Galaxy formation is a complicated combination of  many different physical processes,
such as gravitational collapse (Eggen et al. 1962), 
accretion of subgalactic clumps (e.g., Searle \& Zinn 1978),
hierarchical merging of dark matter halos
(e.g., White \& Rees 1978),
star formation and its feedback effects (e.g., Larson 1974; Dekel \& Silk 1986),
growth of bars and spiral arms (e.g., Sellwood \& Carlberg 1984; Athanassoula 2003),
clump dynamics (e.g., Noguchi 1999),
feedback effects of active galactic nuclei (e.g., Di Matteo et al. 2005;
Springel et al. 2005),
hot and cold mode gaseous accretion  
(e.g., Keres et al. 2005; Dekel \& Birmboim 2006),
and environmental influences of groups and clusters (e.g., Dressler 1980;
Bekki \& Couch 2011).
It has been a key issue for Galactic and extragalactic  astronomy to understand
the relative importance of each of
these physical processes in the formation and evolution
of galaxies with different
masses and Hubble types residing in different environments at a give cosmic time.
In order to address this key issue, many theoretical studies have tried
to construct as realistic galaxy formation models as possible by using
sophisticated numerical simulations.

One of the missing elements in these previous numerical studies
of galaxy formation and evolution is interstellar dust. 
The efficient conversion from neutral to  molecular hydrogen (${\rm H_2}$) 
is possible on the surface of dust grains
(e.g., Gould \& Salpeter 1963; Hollenbach \& Salpeter 1971),
and thermodynamical evolution of ISM and interstellar chemistry
are also controlled by dust
(e.g. Herbst 2002).
Furthermore,
fragmentation processes
of metal-poor gas clouds can be significantly influenced by
dust abundances so that dust can be very important for 
the formation of low-mass stars in
the early universe (e.g., Larson 2005; Schneider \& Omukai 2010).
In spite of these importance of dust,  dust-related physical processes
have not been self-consistently included in almost all simulations of
galaxy formation and evolution. 
The observed physical processes of  dust properties in galaxies
have been investigated so far
by one-zone chemical evolution models
(e.g., Dwek 1998, D98; Lisenfeld \& Ferrara 1998; Hirashita 1999; Edmunds 2001;
Inoue 2003;
Calura et al. 2008; Asano et al. 2013; Zhukovska \& Henning 2013;
Rowlands et al. 2014).

Bekki (2013a, 2015; B13a and B15, respectively) 
have first performed numerical simulations of galaxy formation
and evolution which include self-consistently
both (i)  the formation, growth, and destruction processes
of dust grains in asymptotic giant branch (AGB) stars, supernovae (SNe), and ISM
and (ii) ${\rm H_2}$ formation on dust grains. 
Thanks to the self-consistent modeling of dust physics, 
not only the cosmic evolution
of dust but also the physical properties of galaxies (e.g., gas contents) were compared
with the latest observations (e.g., by Herschel) in B13a and B15.
These simulations are different from previous ones 
(Ferrara et al. 1991, F91; Aguirre et al. 2001) in which dust grains are
represented by
'test particles' (i.e., moving in a gravitational  potential of a galaxy
without influencing its host galaxy  at all)
so that the influences of dust on galaxy formation and evolution
can not be investigated. Although our previous simulations made it possible
for us to predict the cosmic evolution histories of dust abundances and distributions
in galaxies, they have some problems in modeling dust.

One key problem is that gas and dust are assumed to move exactly the same way
in our previous simulations (B13a, B15): 
the simulated spatial distributions and kinematics
are the same between gas and dust in galaxies. 
This could be a good approximation, given that the timescale of gas-dust
frictional timescale is an order of $\sim 10^3$ yr for a typical
range of ISM properties (e.g., Theis \& Orlova 2004), which is much shorter
than the dynamical timescales of most galaxies ($10^7-10^8$ yr).
However, as shown in previous
theoretical works on dust evolution in galaxies,  only dust can be expelled
from galactic disks through the effects of  radiation pressure of stars 
on dust (e.g., F91). Also, the observed flat distribution of
halo dust (M\'enard et al. 2010, M10) and the  rather high dust-to-gas-ratio
($D \sim 0.05$) in  M81 group  (e.g., Xilouris et al. 2006)
strongly suggest that the spatial distributions of dust can be quite different
from those of gas in some galaxy environments.

Furthermore, chemical evolution of galaxies can be significantly influenced by
dust removal from galaxies through radiation-driven winds (Bekki \& Tsujimoto 2014),
which implies that star formation processes in galaxies
can be influenced by dust removal from galaxies
owing to the processes dependent on dust and metal abundances.
Therefore,  we would need to model dust removal processes in galaxies
more self-consistently so that we can discuss the roles of dust in 
galaxy formation and evolution and the observed  wide variety of 
dust properties of galaxies in a more comprehensive way.
It would be a reasonable strategy for numerical studies of galaxy formation
to treat  gas and dust separately, because gas and dust respond differently
to physical processes of ISM.

The purpose of this paper is to describe a new method by which dust-related
physics can be more self-consistently included in Nbody/hydrodynamical
simulations of galaxy formation and evolution. 
Dust and gas are represented by separate 'live' particles 
in the present new simulation code so that dust-related physical processes
such as radiation pressure on dust and dust-gas interaction can be
better investigated.
Since this is the very first paper discussing the possible roles of dust in 
galaxy formation based on a new type of galaxy evolution simulations,
we focus mainly on the methods to include dust-related physics
in Nbody/hydrodynamical simulations. We briefly describe some key results
of the new simulations, and we discuss extensively the possible roles
of dust in galaxy formation and evolution in forthcoming papers.

The present model is not only a more sophisticated  version of previous models on
the formation of dust wind (e.g., Chiao \& Wickamasinghe 1972;
F91) but also an improved version of previous galaxy evolution models
for dusty galaxies. The important influences of radiation-driven dust wind
on star-forming regions in  galaxies have been recently discussed by a number of
authors (e.g., Thompson et al. 2005;  Hopkins et al 2012).
These works, however, did not investigate  the evolution of dust contents
and the possibly separate distributions of gas and stars in a self-consistent
manner. 
Furthermore, previous models for spectral energy distributions (SED) of galaxies
(e.g., Bekki  \& Shioya 2000; Jonsson 2006) 
and for ${\rm H_2}$ contents of galaxies 
(e.g., Pelupessy et al. 2006; Krumholz et al. 2009)
assumed a fixed dust-to-metal ($D_{\rm z}$) ratio across the entire region of
a simulated galaxy. Such a fixed $D_{\rm z}$ is neither observed 
(e.g., Galametz et al. 2011) nor realistic in theoretical modeling
of dust evolution (e.g., Inoue 2003). Therefore, the present model
with variable $D_{\rm z}$ in different local regions of a simulated galaxy
can provide more detailed and accurate predictions on ${\rm H_2}$ contents
and SEDs of galaxies.

The plan of the paper is as follows. First, 
we describe the details of the adopted new methods to implement
dust-related physical processes in galaxy-scale chemodynamical simulations.
Then, we present some key results of the new simulations on the roles
of dust in the evolution of  star formation rates, dust distributions,
and gas dynamics  within galaxies in \S 3.
In this section, we also discuss how the present results depend on
the physical parameters of the adopted new method for dust-related physics.
In \S 4, we discuss (i) the advantages and disadvantages of the present new method
over previous ones and (ii) further improvement of the method in our future works.
We summarize our  conclusions in \S 5.

Although including dust-related physics in galaxy-scale Nbody+hydrodynamical simulations
is rather new in numerical studies of galaxy formation and evolution, 
there are already a number of more sophisticated simulations on
the joint evolution of dust and gas 
in other areas of astronomy. For example, Laibe \& Price (2014)
have already included the gas-dust interaction for multiple dust components in their
SPH simulations that can be used for the investigation of planet formation and evolution.
The methods and numerical techniques used in these simulations in other areas of
astronomy would be useful in the present study. However, we do not discuss these,
because this is beyond the scope of this paper. We accordingly focus on the
modeling issues of galaxy-scale chemodynamical simulations with dust physics.

\begin{table*}
\centering
\begin{minipage}{175mm}
\caption{
A list of physical processes implemented in
the simulations codes used in B13a and this work.
This table accordingly shows a number  key  differences between the two codes.
}
\begin{tabular}{lccl}
{ Physical effects }
& {B13a \footnote{ $\bigcirc$ and $\times$  in the second
column  mean
inclusion and non-inclusion of the listed physical effect, respectively}}
& This work 
& Specifications \\
Dust formation  & $\bigcirc$ & $\bigcirc$ &
Formation in SNe and AGB stars \\
Dust destruction  & $\bigcirc$ & $\bigcirc$ &
Destruction by  SNe  \\
Dust shattering in halo gas & $\times$ & $\times$ & \\
Dust coagulation  & $\times$ & $\times$ & \\
Dust growth & $\bigcirc$ & $\bigcirc$  &
Variable dust accretion timescale
in this work\\
Size evolution of dust  & $\times$ & $\times$ &  \\
Size-dependence on dust composition  & $\times$ &  $\bigcirc$ & 
Only two components (silicate and graphite) \\
Stellar radiation pressure on dust   & $\times$ & $\bigcirc$ & \\
Gas-dust hydrodynamical coupling   & $\times$ & $\bigcirc$ & \\
Photo-electric heating   & $\times$ & $\bigcirc$ & \\
Gas-dust heating   & $\times$ & $\bigcirc$ & \\
Cosmic-ray heating   & $\times$ & $\bigcirc$ & \\
UV background radiation   & $\times$ & $\times$ &  \\
${\rm H_2}$ formation on dust grains & $\bigcirc$ &  $\bigcirc$ & \\
${\rm H_2}$ formation efficiency dependent on $D$ and $ISRF$  & $\bigcirc$ &  $\bigcirc$ & \\
${\rm H_2}$ photo-dissociation by ISRF  & $\bigcirc$ & $\bigcirc$  &
ISRF locally defined for each gas particle \\
Star formation & $\bigcirc$ & $\bigcirc$ &
${\rm H_2}$-dependent recipe \\
SN feedback effects  & $\bigcirc$ & $\bigcirc$ &
Both SNII and prompt SNIa are included.\\
AGN feedback effects  & $\times$ &  $\times$ & \\
Growth of SMBHs   & $\times$ & $\times$ & \\
Chemical evolution  & $\bigcirc$ & $\times$ &
11 elements (e.g., C, N, and O) \\
Chemical enrichment by AGB ejecta & $\bigcirc$ & $\bigcirc$ & \\
Dust-corrected cooling   & $\bigcirc$ & $\bigcirc$ &
[Fe/H] corrected by dust depletion is used.  \\
Metallicity-dependent radiative cooling   & $\bigcirc$ & $\bigcirc$ &
For $T_{\rm g}>10^4$ K  \\
${\rm [CII]+[OI]}$ cooling  & $\times$ & $\bigcirc$ & Different cooling
rate adopted in B13a\\
${\rm H_2}$ cooling   & $\times$ & $\bigcirc$ &
For $T_{\rm g}>10$ K  \\
\end{tabular}
\end{minipage}
\end{table*}

\section{The model}

\subsection{A new four-component galaxy model}
The most remarkable difference in dust modeling between B13a and the present study
is that gas and dust  in ISM of galaxies are represented by separate 'live' particles.
Accordingly, a galaxy consists of dark matter, stars, gas, and dust 
('four-component' model) in the present
study, which is different from almost all previous simulations of galaxy
formation and evolution in which
a galaxy consists of dark matter, gas, and stars (`three-component').
Dust particles gravitationally interact both with other components
of galaxies (e.g. dark matter and stars) and with other dust particles
in the new four-component model.  Furthermore,
gas-dust interaction and response of dust to radiation fields of stars can be
more self-consistently included in the four-component  model. As in B13a,
${\rm H_2}$ formation on dust grains and the formation, growth, and destruction
processes of dust due to star formation and SN feedback effects are included
in the present model. 
By using an isolated disk galaxy model described in this
section, we mainly discuss how radiation pressure of stars on dust grains
can influence ${\rm H_2}$ formation processes and star formation histories of
galaxies.

The isolated disk model adopted in this study is the same as those in Bekki (2014b)
and therefore we briefly describe it here. The models for ${\rm H_2}$ formation
on dust grains and the formation, growth, and destruction processes of dust in ISM
of galaxies are essentially the same as those adopted in B13a. We thus describe
the details of the models only when they are differently implemented in the present
new simulations with live dust particles. Some new ingredients that were not
included in B13a are described and discussed in detail in this section.
Dust-related physical processes that are missing in the present model
yet could be important in ISM evolution are discussed later in \S 4.

The new simulation code used in this work is based on the one
developed in our previous works (B13a, B15)
that  can be run on clusters of GPU (Graphics Processing Unit) machines.
In the new code,
gravitational
calculations of many particles
can be done on GPUs whereas other calculations (e.g., star formation
and hydrodynamics) can be done on CPUs.
The GPUs are used also in calculating the radiative force
(due to radiation pressure of a star particle on a dust particle)  which is 
proportional to $L_{\rm star}/r^2$, where $L_{\rm star}$ is
the total stellar luminosity and $r$ is the distance between the dust and
star particles.
The code adopts
the smoothed-particle hydrodynamics (SPH) method for following the time
evolution of gas dynamics in galaxies. 
The key differences in the adopted models between B13a and this work
are summarized in Table 1 and the physical meanings for symbols used
in this work are given in  Table 2.

\subsection{Disk galaxy model}

In the present disk model,  a galaxy consists of dark matter, stars, gas,
and dust: this is the 'four-component' disk model as opposed to the `three-component'
(i.e., dark matter + stars + gas)
one adopted in almost all previous simulations of galaxy evolution.
The total masses of dark matter halo, stellar disk, gas disk,
bulge, and dust  of a disk galaxy are denoted as $M_{\rm h}$, $M_{\rm s}$, $M_{\rm g}$,
$M_{\rm b}$, and $M_{\rm dust}$, respectively. 
In order to describe the initial density profile of dark matter halo
in a disk galaxy,
we adopt the density distribution of the NFW
halo (Navarro, Frenk \& White 1996) suggested from CDM simulations:
\begin{equation}
{\rho}(r)=\frac{\rho_{0}}{(r/r_{\rm s})(1+r/r_{\rm s})^2},
\end{equation}
where  $r$, $\rho_{0}$, and $r_{\rm s}$ are
the spherical radius,  the characteristic  density of a dark halo,  and the
scale
length of the halo, respectively.
The $c$-parameter ($c=r_{\rm vir}/r_{\rm s}$, where $r_{\rm vir}$ is the virial
radius of a dark matter halo) and $r_{\rm vir}$ are chosen appropriately
for a given dark halo mass ($M_{\rm dm}$)
by using the $c-M_{\rm h}$ relation
predicted by recent cosmological simulations
(e.g., Neto et al. 2007).

The bulge of a disk galaxy  has a size of $R_{\rm b}$
and a scale-length of $R_{\rm 0, b}$
and is represented by the Hernquist
density profile. The bulge is assumed to have isotropic velocity dispersion
and the radial velocity dispersion is given according to the Jeans equation
for a spherical system.
The radial ($R$) and vertical ($Z$) density profiles of the stellar disk are
assumed to be proportional to $\exp (-R/R_{0}) $ with scale
length $R_{0} = 0.2R_{\rm s}$  and to ${\rm sech}^2 (Z/Z_{0})$ with scale
length $Z_{0} = 0.04R_{\rm s}$, respectively.
The gas disk with a size  $R_{\rm g}=2R_{\rm s}$
has the  radial and vertical scale lengths
of $0.2R_{\rm g}$ and $0.02R_{\rm g}$, respectively.
In addition to the
rotational velocity caused by the gravitational field of disk,
bulge, and dark halo components, the initial radial and azimuthal
velocity dispersions are assigned to the disc component according to
the epicyclic theory with Toomre's parameter $Q$ = 1.5.
The vertical velocity dispersion at a given radius is set to be 0.5
times as large as the radial velocity dispersion at that point.

\subsection{Live dust particle}

The initial distribution of dust and kinematics
in a disk galaxy are  assumed to be the same as
those of the gas. The `dust disk' of a galaxy 
is represented by live dust particles
with the total particle number of  $N_{\rm dust}$: this $N_{\rm dust}$ can
increase according to the star formation history of the galaxy.
The evolution of each dust particle can be influenced 
by (i) gravitational force from other components
(e.g., stars) and other dust particles ($F_{\rm grav}$), (ii) radiation pressure
of stars ($F_{\rm rad}$), and (iii) drag effects of surrounding gas particles
($F_{\rm drag}$).
Thus, the total force exerting on the $i$-th dust particle is as follows:
\begin{equation}
F_i=F_{\rm grav,\it i}+F_{\rm rad, \it i}+F_{\rm drag, \it i}.
\end{equation}
The dust particles can be destroyed by SNe so that their masses ($m_{\rm d}$)  can be drastically
reduced when star formation occurs around the particles.
It should be noted here that 
this dust particle mass ($m_{\rm d}$) is different from dust grain mass ($m_{\rm dust}$)
for each dust component (silicate or graphite),
which is later used. The dust particle is assumed to consist of numerous dust grains
(e.g., silicate and graphite), and accordingly, $m_{\rm d}=\eta_{\rm dust} m_{\rm dust}$,
where $\eta_{\rm dust}$ corresponds to the number of dust grains in a dust particle.

Since gas and dust  move independently from each other in the present simulations,
the spatial distributions of gas and dust in a galaxy  can be  different.
Therefore, we estimate the ${\rm H_2}$ formation on dust grains at each dust 
particle's position at each time step. This is quite different from B13a in
which ${\rm H_2}$ formation is estimated at each gas particle's position
owing to the adopted assumption that gas and dust always have the same spatial
distribution.
The growth and destruction processes of dust are also estimated at each
dust particle's position accordingly to the physical properties of gas around
the dust particle. The detailed methods to implement these are described
later in this section. The dependence of the present results on $N_{\rm dust}$
is briefly discussed in Appendix A.

\begin{table}
\centering
\begin{minipage}{80mm}
\caption{Description of physical meanings for symbols often used in the present study.}
\begin{tabular}{ll}
{Symbol}
& {Physical meaning}\\
RP  &  Radiation pressure of stars on dust grains  \\
PEH  &  Photo-electric heating of gas by dust \\
GD  &  Gas-dust heating of ISM \\
CR  &  Cosmic-ray  heating of ISM \\
DR  &  Gaseous drag of dust  \\
$D$   &  dust-to-gas ratio \\
$f_{\rm H_2}$ & mass fraction of molecular hydrogen (${\rm H_2}$) \\
$Q_{\rm pr}^{\ast}$   &  Frequency-averaged radiation pressure coefficient \\
$A_{\rm t}$   &  Total dust extinction (mag) \\
$t_{\rm 0}$   &  Initial ages of stars in  a disk \\
$f_{\rm dest}$   &  The mass fraction dust destroyed by SNe \\
\end{tabular}
\end{minipage}
\end{table}

\subsection{Gravitational dynamics and hydrodynamics}

Since the total number of particles ($N$) used in this study is an order of $10^6$,
we adopt a direct-summation N-body algorithm 
for gravitational interaction
between dark matter, stars, gas, and dust.
The calculation speed of the direct summation method
can be significantly increased  by using GPUs or GRAPE-DR, which
is the latest version of the special-purpose computer for gravitational dynamics.
In the present code,
the gravitational softening length ($\epsilon$) is chosen for each
component in a galaxy (i.e.,
multiple gravitational softening lengths).
Thus the gravitational softening length ($\epsilon$)
{\it can be}  different between dark matter ($\epsilon_{\rm dm}$), stars
($\epsilon_{\rm s}$), 
gas ($\epsilon_{\rm g}$), and dust ($\epsilon_{\rm d}$),
and $\epsilon$ is determined by the initial
mean separation of each component.

The gravitational softening lengths for stars, gas, new stars,
and dust are set to be the same with one another.
Initial $\epsilon_{\rm g}$ is set to be significantly smaller than
$\epsilon_{\rm dm}$ owing to rather  high number-density of gas particles.
Furthermore,  when two different components interact gravitationally,
the mean softening length for the two components
is applied for the gravitational calculation.
For example, $\epsilon = ({\epsilon}_{\rm dm}+{\epsilon}_{\rm g})/2$
is used for gravitational interaction between gas and dark matter.
The values of $\epsilon_{\rm dm}$ and $\epsilon_{\rm g}$ ($=\epsilon_{\rm s}$)
are 2.1 kpc and 0.2 kpc, respectively, for the fiducial MW-type disk model
later described.
Thus, the gravitational softening length for interaction between $i$th and $j$th
particles is as follows:
\begin{equation}
\epsilon_{i,j}=\frac{ \epsilon_i + \epsilon_j} {2},
\end{equation}
where $\epsilon_i$ and $\epsilon_j$ take either of the above five 
(virtually two) values
for the five components.

We consider that galactic ISM composed of gas and dust
 can be modeled as an ideal gas with
the ratio of specific heats ($\gamma$) being 5/3.
The gaseous temperature ($T_{\rm g}$) is set to be $10^4$ K initially in all models.
The basic methods to implement SPH in the present study are essentially
the same as those proposed by Hernquist \& Katz (1989).
We adopt the predictor-corrector algorithm (that is accurate to second order
in time and space) in order to integrate the equations
describing  the time  evolution of a system.
Each particle is allocated an individual time step width 
($\Delta t$) that is determined
by physical properties of the particle.
The maximum time step width ($\Delta t_{\rm max}$)
is $0.01$ in simulation units, 
which means that  $\Delta t_{\rm max}=1.41 \times 10^6$ yr
in the present study. 
Star formation rates, chemical abundances, ${\rm H_2}$ fractions,  and dust properties
are all updated once for  every $\Delta t_{\rm max}$.

The radiative cooling processes
for $T_{\rm g} > 10^4$K
are properly included  
by using the cooling curve by
the MAPPINGS III code
(Sutherland \& Dopita 1993).
It should be stressed here that in estimating the cooling rate
depending on gas-phase metallicity and temperature,
we use the correct gas-phase metallicity for each gas gas particle.
Here 'correct' means that dust-phase metals (i.e., those locked up into
dust grains thus can not participate radiative cooling) are excluded
from the estimation of gas-phase metallicities. 
In estimating gas-phase [Fe/H] in B13a and B15, 
we calculate the gas-phase Fe mass ($m_{\rm Fe,g}$) 
for each particle at each time step as follows:
\begin{equation}
m_{\rm Fe,g}=m_{\rm Fe,t}-m_{\rm Fe,d},
\end{equation}
where $m_{\rm Fe, t}$ and $m_{\rm Fe,d}$ are the total Fe 
mass and Fe mass locked up in dust for the particle. 
In the present model, we can simply use the total Fe  mass of 
a gas particle as $m_{\rm Fe,g}$, because 
the evolution of gas and that of dust are separately examined at each time step
(i.e., $m_{\rm Fe, d}$ is always 0 for gas).
Thus, the present
simulation can better estimate the cooling rates for gas particles than
previous simulations that do not include dust.

For  $10 \le T_{\rm g} \le 10^4$K, we 
adopt two analytic formulae proposed by Wolfire et al. (2003) for gas cooling
in neutral atomic regions (e.g., by [CII] 158$\mu$m cooling) and
by Galli \& Palla (1998) for ${\rm H_2}$ cooling.
The cooling rate coefficient by Wolfire et al. (2003) is as follows:
\begin{equation}
\Lambda_{\rm CII+OI} = 5.4 \times 10^{-27} T_2^{0.2}
e^{-1.5/T_2} Z_{\rm g}^{'} \; {\rm ergs}  \; {\rm cm^3} \; {\rm  s^{-1} },
\end{equation}
where $T_2=T_{\rm g}/100$ ($T_{\rm g}$ is the gas temperature in units of K)
and $Z_{\rm g}^{'}$ is the gas-phase metallicity normalized by the solar
value (0.02). 
Since this cooling rate is per $n_{\rm H}^2$ (where $n_{\rm H}$ is the hydrogen
number density),  $n_{\rm H}$ needs to be calculated for each SPH particle at
each time step for estimating the cooling rate.
The above adopted formula is an approximation that
Wolfire et al. (2003)  derived
from full numerical calculations of ISM by assuming that
H atoms are responsible for the excitation of [CII] and [OI] for
the gas density less than $3000$ cm$^{-3}$ and the gas temperature
ranging from 100 K to 1000 K.

We adopt the equation A7 described in Galli \& Palla (1998) for ${\rm H_2}$ cooling
and the formula is as follows:
\begin{eqnarray}
\log \Lambda_{\rm H_2} =-103.0 +97.59\log T_{\rm g}-48.05(\log T_{\rm g})^2 
\nonumber \\
+10.80(\log T_{\rm g})^3-0.9032(\log T_{\rm g})^4 .
\end{eqnarray}
Since this cooling rate is per $n_{\rm H} n_{\rm H_2}$,
we need to calculate
the hydrogen number density ($n_{\rm H}$) and ${\rm H_2}$ number density
($n_{\rm H_2}$) of a SPH gas particle in order to estimate
the ${\rm H_2}$ cooling rate for the  particle.
In the above ${\rm H_2}$ cooling formula, galaxies with higher dust abundances
are likely to show more efficient ${\rm H_2}$ cooling, because 
${\rm H_2}$ formation rates are higher in such galaxies.
Therefore, thermal evolution of ISM and thus gas dynamics of galaxies can be 
significantly influenced by dust evolution in the present study.

As shown in Galli \& Palla (1998) and Hollenbach \& McKee (1989),
this ${\rm H_2}$ cooling rate depends on gas density too. Therefore, the adopted
formula with a temperature-dependence only could be less realistic for 
describing real ISM evolution influenced by ${\rm H_2}$-related physical processes. 
For example,
${\rm H_2}$ cooling is much less efficient in high-density ISM
with $n_{\rm H} \sim 10^6$ cm$^{-3}$, as demonstrated
in Figure A1 of Galli \& Palla (1998). Although such high-density gaseous regions
can be rarely formed owing to the adopted resolution in the present simulations,
the model without density-dependent ${\rm H_2}$ cooling
would mean that (i) the present simulations
overestimate the net cooling of gas in disk galaxies
and (ii) gas clumps can be more efficiently formed and thus star formation
can be more efficient.

The influences of the UV background radiation is not taken into account in the
present models either, which means that heating of ISM by the UV 
background radiation (and  internal UV radiation from young massive
stars and active galactic nuclei, AGN, in a simulated galaxy)
is not properly modeled. This heating of ISM could be more important for 
the gas dynamical evolution of low-mass
disk galaxies where radiative cooling is less efficient owing to lower metallicities.
This possible important effect will need to be properly included in our future
galaxy-scale simulations with dust-related physical processes.

\subsection{Star formation}

A gas particle is converted into a `new star' (collisionless particle)
if the following three SF conditions (i)-(iii) are satisfied:
(i) the local dynamical time scale is shorter
than the sound crossing time scale (mimicking
the Jeans instability) , (ii) the local velocity
field is identified as being consistent with gravitationally collapsing
(i.e., div {\bf v}$<0$),
and (iii) the local density exceeds a threshold density for 
star formation ($\rho_{\rm th}$).
This check of gas-to-star-conversion is done at every $0.01 t_{\rm unit}$
(corresponding to the maximum time step width, $\Delta t_{\rm max}$),
where $t_{\rm unit}$ is the time units ($1.41 \times 10^8$ yr)
adopted in the present simulations. In our previous simulations,
gas-to-star-conversion is done whenever the above three
conditions are satisfied. Therefore, the present way to model star formation
is a bit different from those in our previous ones.

Star formation can increase dramatically the total number of new stars
and dust, if only some fractions of gas particles continue
to be converted into new stars
at each time step until the mass of the gas particle
becomes very small. 
Since this method is  numerically very costly and practically infeasible,
we adopt the following SF conversion method.
A  gas (for which the above three SF conditions are satisfied)
is regarded as having  a SF probability ($P_{\rm sf}$);
\begin{equation}
P_{\rm sf}=1-\exp ( -C_{\rm eff} f_{\rm H_2}
\Delta t_{\rm max} {\rho}^{\alpha_{\rm sf}-1} ),
\end{equation}
where $C_{\rm eff}$ corresponds to a star formation  efficiency (SFE)
in molecular cores and is set to be 0.3.
$f_{\rm H_2}$ is the ${\rm H_2}$ mass fraction of the gas particle,
$\Delta t_{\rm max}$ is the maximum time step width for the particle,
$\rho$ is the gas density of the particle,
and $\alpha_{\rm sf}$ is
the power-law slope of the  Kennicutt-Schmidt law
(SFR$\propto \rho_{\rm g}^{\alpha_{\rm sf}}$;  Kennicutt 1998).
A reasonable value of
$\alpha_{\rm sf}=1.5$ is adopted in the present
study.
This SF probability has been already introduced in our early chemodynamical
simulations of galaxies (e.g., Bekki \& Shioya 1998).
By generating random numbers, we implement this SF method (See B13a and B15
for this).

Each SN is assumed to eject the feedback energy ($E_{\rm sn}$)
of $10^{51}$ erg and 90\% and 10\% of $E_{\rm sn}$ are used for the increase
of thermal energy (`thermal feedback')
and random motion (`kinematic feedback'), respectively.
The thermal energy is used for the `adiabatic expansion phase', where each SN can remain
adiabatic for a timescale of $t_{\rm adi}$.
This adiabatic model is adopted both for SNIa and SNII.
Although $t_{\rm adi}=10^5$ yr is reasonable for a single SN explosion,
we adopt a much longer $t_{\rm adi}$ of $\sim 10^6$ yr.
This is mainly because multiple  SN explosions can occur for a gas particle with a mass of
$10^5 {\rm M}_{\odot}$ in these galaxy-scale simulations,
and $t_{\rm adi}$ can be different
for multiple SN explosions in a small local region owing to complicated
interaction between gaseous ejecta from different SNe.
Such interaction of multiple  SN explosions would make the adiabatic phase
significantly longer in real ISM of galaxies.
We adopt a canonical  Salpeter stellar initial mass function (IMF)
with the slope ($\alpha_{\rm IMF})$ of $-2.35$ 
and the upper and lower cutoff masses being $0.1 {\rm M}_{\odot}$ and
$100 {\rm M}_{\odot}$, respectively,
for all models in the present study.

\begin{table*}
\centering
\begin{minipage}{160mm}
\caption{Description of the basic parameter values
for the three different disk galaxy  models.}
\begin{tabular}{lcccc}
Model name/Physical properties
& MW-type  & LMC-type  & Dwarf-type & Sa-type \\
DM mass ($\times 10^{12} {\rm M}_{\odot}$) & 1.0 & 0.1 & 0.01 & 1.0 \\
Virial radius  (kpc) & 245.0  & 113.8 & 52.9 & 245.0 \\
{$c$  \footnote{$c$ is the $c$-parameter in the NFW dark matter
profiles.}}
&  10 & 12 & 16 & 10 \\
Stellar disk  mass ($\times 10^{10} {\rm M}_{\odot}$) & 6.0 & 0.36 & 0.036 
& 6.0 \\
Gas disk  mass ($\times 10^{10} {\rm M}_{\odot}$) & 0.6 & 0.18 & 0.0036 & 0.6 \\
Bulge  mass ($\times 10^{10} {\rm M}_{\odot}$) & 1.0 & -- & -- & 12.0 \\
{Dust-to-gas-ratio \footnote{The value is adopted from
observational results by Zubko et al. 2004.}}  & 0.0064 & 0.0064 & 0.0064 & 0.0064 \\
Dust-to-metal-ratio & 0.4 & 0.4 & 0.4 & 0.4\\
Initial central 
gas-phase metallicity ([Fe/H]$_0$)  & 0.34 & $-0.28$ & $-1.04$ & 0.34\\
Radial metallicity gradient (dex kpc$^{-1}$)  & $-0.04$ & $-0.04$ & $-0.04$
& $-0.04$  \\
Stellar disk size (kpc) & 17.5 & 8.1 & 3.8 & 17.5\\
Gas disk size (kpc)  & 35.0 & 16.2 & 7.6 & 35.0\\
Bulge  size (kpc) & 3.5 & -- & -- & 17.2\\
Gas disk  size (kpc) & 35.0 & 16.2 & 7.6 & 35.0 \\
DM particle mass ($\times 10^6 {\rm M}_{\odot}$)  & 
1.4 & 0.14 & 0.014 & 1.4 \\
Gas particle mass ($\times 10^4 {\rm M}_{\odot}$)  & 
6.0 & 1.8 & 0.36 & 6.0 \\
DM softening length ($\epsilon_{\rm DM}$)  & 2.1 kpc  & 0.6 kpc  & 0.17 kpc 
& 2.1 kpc\\
Gas softening length ($\epsilon_{\rm g}$)  & 0.2 kpc  & 0.063 kpc & 0.02 kpc 
& 0.2 kpc\\
\end{tabular}
\end{minipage}
\end{table*}

\subsection{Chemical enrichment}

The present model for chemical enrichment processes of galaxies
is essentially similar to those  used in B13a and B15,
though the minor details are different between these works.
Therefore, we briefly describe only the key elements of the
model here.
Chemical enrichment through star formation and metal ejection from
SNIa, II, and AGB stars is self-consistently included in the chemodynamical
simulations.
We investigate the time evolution of the 11 chemical elements of H, He, C, N, O, Fe,
Mg, Ca, Si, S, and Ba in order to predict both chemical abundances and dust properties
in the present study.
We consider the time delay between the epoch of star formation
and those  of supernova explosions and commencement of AGB phases (i.e.,
non-instantaneous recycling of chemical elements).

We adopt the `prompt SN Ia' model in which
the delay time distribution (DTD)
of SNe Ia is consistent with  recent observational results by  extensive SN Ia surveys
(e.g.,  Mannucci et al. 2006).
We adopt the nucleosynthesis yields of SNe II and Ia from Tsujimoto et al. (1995; T95)
and AGB stars from van den Hoek \& Groenewegen (1997; VG97)
in order to estimate chemical yields in the present study.
In the present study,  ejecta from SNe and AGB stars can become either dust 
particles or gas-phase metals in gas particles. Therefore, the implementation
of chemical enrichment process in the present study with live dust particles
is different from those adopted in B13a and B15.

Metals ejected from a stellar particle (SNIa or SNII or AGB star) are assumed
to be equally distributed among neighboring SPH gas particles around the stellar particle.
These `neighboring particles' are selected for each SN or AGB star,
if SPH particles satisfy the following condition;
\begin{equation}
R \le R_{\rm mix},
\end{equation}
where $R$ is the distance between a SPH particle and a SN (or AGB star)
and $R_{\rm mix}$ is the 'mixing radius' (parameter).
The introduced mixing radius ($R_{\rm mix}$) is set to be the same as the
softening length for gas from the stellar particle. Although metals can be locally
mixed with gas in this  model, the mixing process of metals in real ISM  might be
more complicated than homogeneous mixing  described in the present  model.
The particle-to-particle abundance differences (corresponding to abundance differences
in different local gas clouds) therefore might not be so realistically modeled in the
present study.

A new stellar particle ('new star'), which is regarded as 
a cluster of numerous stars with a given IMF,  eject dust and metals only
three times: $2.7\times 10^7$ yr (corresponding to SNII explosions,
denoted as $t_{\rm SNII}$),
$10^8$ yr (prompt SNIa explosions, $t_{\rm SNIa}$),
and $2.3 \times 10^8$ (onset of AGB phases, $t_{\rm AGB}$)
after the star formation. 
A dust particle with chemical elements derived
from the adopted IMF and dust yield is created 
at each ejection process. 
Therefore, the total  mass of a star born at the time $t=t_{\rm form}$
can reduce at each ejection process ($t_{\rm form}<t$).
For example, the total mass of  $j$-th element for  $i$-th stellar particle
($m_{\rm s, \it i, j}$)
after SNII explosions is as follows. 
\begin{eqnarray}
m_{\rm s, \it i,j}(t=t_{\rm form}+t_{\rm SNII}) = 
m_{\rm s,\it i,j}(t=t_{\rm form}) 
\nonumber \\
- m_{\rm d, \it i,j} -
\Delta m_{\rm ej, \it i, j},
\end{eqnarray}
where $m_{\rm d,\it i,j}$ is the total mass of  $j$-th element
of a dust particle produced by  $i$-th stellar particle
and $\Delta m_{\rm ej,\it i,j}$ is the total mass of  $j$-th element
of the ejected gas from  $i$-th stellar particle.
The total mass of $j$-th element of  $k$-th gas particle
around  $i$-th stellar particle increases as a result
of metal ejection as follows:
\begin{eqnarray}
m_{\rm g, \it k,j}(t=t_{\rm form}+t_{\rm SNII}) = 
m_{\rm g,\it k,j}(t=t_{\rm form}) 
\nonumber \\
+\Delta m_{\rm ej, \it i, j}/N_{\rm nei, \it i},
\end{eqnarray}
where $N_{\rm nei, \it i}$ is the total number of gas particles
around the stellar particle (i.e., within the SPH-smoothing length,
$h_k$, for $k$-th gas particle).
By replacing $t_{\rm SNII}$ with $t_{\rm SNIa}$ and $t_{\rm AGB}$,
we can estimate the mass evolution of stellar and gaseous particles due to
SNIa explosions and stellar mass loss of  AGB stars.

We limit the number of
dust and metal ejection to only three times,
because we need to avoid an unnecessarily large number of particles that
can virtually make it impossible for us to simulate galaxy evolution
owing to the dust-related calculations.
The dust particles initially in thin disks
and  those from SNIa, SNII, and AGB stars
are labeled as `old', `SNIa', `SNII', and `AGB' dust, respectively.
The maximum total number of particles of a  simulation
for which we can finish all calculations of a model within a reasonable timescale
(less than a week)  on
a single GPU server (Tesla K10 used in this paper) is around $2 \times 10^6$.
This means that we need to have a massively parallel cluster to run
$N \sim 10^8-10^9$ for cosmological hydrodynamical simulations with dust dynamics
and recycling. We discuss how to implement the present new code to run
on such a large cluster in our forthcoming papers.

The initial chemical abundances and dust properties are different in different
gas particles and given according to the positions of the particles within
its host disk galaxy.
The gas-phase  metallicity of each (gaseous and stellar) particle is given
according to its initial position:
at $r$ = $R$,
where $r$ ($R$) is the projected distance (in units of kpc)
from the center of the disk, the metallicity of the star is given as:
\begin{equation}
{\rm [Fe/H]}_{\rm r=R} = {\rm [Fe/H]}_{\rm d, r=0} + \alpha \times {\rm R}. \,
\end{equation}
where  $\alpha$ is the slope of the metallicity gradient in units of dex kpc$^{-1}$.
We show the results of the models with $\alpha=-0.04$,
which is  the observed  value  of our Milky Way
(e.g., Andrievsky et al. 2004).Initial dust-to-metal-ratio is set to be 0.4
for all particles in a simulation. The central metallicity 
${\rm [Fe/H]}_{\rm d, r=0}$ is simple referred to as [Fe/H]$_0$ and it is
different in different galaxy models with different masses.

\subsection{Dust growth and destruction}

Although the dust model adopted in the present study is
similar to those adopted by B13a 
and B15,  there are  some differences between these models
owing to the newly adopted live dust particle method.
We here describe the model with special emphasis on the model differences.
We calculate the  total mass of $j$th component ($j$=C, O, Mg, Si, S, Ca, and Fe)
of dust from $k$th type of stars ($k$ = I, II, and AGB for SNe Ia, SNe II, and
AGB stars, respectively) based on the methods described in B13a that is similar to
those adopted in D98.
We consider
that the key parameter in dust accretion is the dust accretion timescale ($\tau_{\rm a}$).
In the present study, this parameter can vary between different gas particles
and is thus represented by $\tau_{\rm a, \it l}$ for  $l$th gas particle.

The mass of $j$th component
($j$=C, O, Mg, Si, S, Ca, and Fe) of $i$th dust particle
at time $t$ can increase owing  to metal accretion
onto the dust from the surrounding gas particles.
The  mass increase of  $j$th element of  $i$th dust particle
due to the metal-transfer between gas and dust particles
is described as
\begin{equation}
\Delta m_{\rm d, \it i,j}^{\rm acc}(t)= 
\Delta t_i 
m_{\rm g, mean, \it i,j}(t) /\tau_{\rm a, mean, \it i},
\end{equation}
where $\Delta t_i$ is the individual time step width for $i$th dust particle
and $m_{\rm g, mean, \it i, j}$ is the mean mass of 
$j$th chemical element for all gas particles around $i$th dust particle,
and $\tau_{\rm a, mean, \it i}$ is the mean dust accretion timescale for the
gas particles.
In order to estimate $m_{\rm g, mean, \it i, j}$ and $\tau_{\rm a, mean, \it i}$,
we choose  a gas particle as a surrounding particle around $i$th dust particle
if it satisfies the following condition:
\begin{equation}
R \le R_{\rm grow},
\end{equation}
where $R$ is the distance between the gas particle and $i$th dust particle
and $R_{\rm grow}$ is a parameter for this neighbor particle search.
Since dust particles do not have smoothing length like SPH particles,
we have to introduce  this parameter
$R_{\rm grow}$.
This $R_{\rm grow}$ is set to be $0.5 \epsilon_{\rm g}$ 
(where $\epsilon_{\rm g}$ is the gravitational length for gas),
and the dependence of the present results on $R_{\rm grow}$ is discussed in Appendix B.

Therefore, $m_{\rm g, mean, \it i, j}$ and $\tau_{\rm a, mean, \it i}$
are estimated as follows: 
\begin{equation}
m_{\rm g, mean, \it i, j} =\frac{1}{N_{\rm nei, \it i} }
\sum_{k=1}^{ N_{\rm nei, \it i} }
m_{\rm g, \it k, j},
\end{equation}
and
\begin{equation}
\tau_{\rm a, mean, \it i} =\frac{1}{N_{\rm nei, \it i} }
\sum_{k=1}^{ N_{\rm nei, \it i} }
\tau_{\rm a, \it k} ,
\end{equation}
where $\tau_{\rm a, \it k}$ is the dust growth timescale 
for $k$th gas particle.
These ways to estimate dust growth become different  from
our previous ones (the equation 9)
in B13a because of the adopted live dust particle method.
Owing to this dust growth, the mass of $j$th chemical
component of $k$th gas particle is reduce and this reduction
is estimated at each time step.

Dust grains can be destroyed though supernova blast waves
in the ISM of galaxies (e.g., McKee 1989)
and the destruction process is parameterized by the destruction time scale
($\tau_{\rm d}$) in previous one-zone models (e.g., Lisenfeld \& Ferrara 1998;
Hirashita 1999). 
Although B13a and B15 adopted a model similar to these one-zone models,
the present study adopts a different one.
The decrease  of the mass of $j$th component
of $i$th dust particle
at time $t$ due to dust destruction process
is as follows
\begin{equation}
\Delta m_{\rm d, \it i,j}^{\rm dest}(t)= - 
f_{\rm dest, \it i} m_{\rm d, \it i,j}(t) 
\end{equation}
where $f_{\rm dest, \it}$ is the mass fraction of dust destroyed by
SNe and assumed to be a free parameter  in the present study.
The possibly reasonable range of this parameter will be discussed later
in this paper.
Although $f_{\rm dust}$ can be different depending on the physical conditions
of ISM (e.g., Jones et al. 1995), we adopted a simplified assumption
of a fixed $f_{\rm dest}$ in this study.
The dust destroyed by SNe can be returned back to the ISM,
and therefore $ f_{\rm dest, \it i} m_{\rm d, \it i,j}(t)/N_{\rm nei, \it i}$ is
added to each of the neighboring gas particle around the dust particle.
This mass increase and decrease is estimated only when SNII and SNIa explosions
occur for each stellar particle. 
This metal-transfer calculation is done  for gas particles, if the particles
satisfy the following condition;
\begin{equation}
R \le R_{\rm dest},
\end{equation}
where $R$ is the distance between a SPH particle and a SN 
and $R_{\rm dest}$ is the 'destruction radius'. This destruction radius
is set to be the same as the mixing radius ($R_{\rm mix}$) for SNe.

Thus the equation for the time evolution of $j$th component of metals
for $i$th dust particle is given as
\begin{equation}
m_{\rm d, \it i,j}(t+\Delta t_i)
=m_{\rm d, \it i,j}(t)+\Delta m_{\rm d, \it i,j}^{\rm acc}(t)
+\Delta m_{\rm d, \it i,j}^{\rm dest}(t)
\end{equation}
When star formation occurs in a molecular cloud,  dust can be consumed by
star formation and locked up in planets around stars. In the present study,
we do not consider this dust consumption by star and planet formation,
mainly because we need to introduce another sets of model parameters.
The total mass of dust locked up in stars and planets could be significantly
smaller than the total amount of dust grown from ISM and produced by stars.
We discuss this important issue in our forthcoming papers.

The dust accretion time scale, $\tau_a$, is 
assumed to be different between different particles
with different gaseous properties and changes with time according to the changes
of gaseous properties in the present study.
In adopting this variable dust accretion ('VDA') model,
we introduce a few additional parameters in VDA in order to
describe the possible dependences of $\tau_{\rm a}$  of gas particles
on the gas densities,
temperature, and chemical abundances.
Our previous simulations with VDA (B15)
have already demonstrated
the importance of dust accretion and destruction in the evolution
of dust-to-gas-ratios ($D$) and ${\rm H_2}$ mass fractions ($f_{\rm H_2}$).
Since the details of the VDA model is given in (B15),
we describe only the key ingredients  of  the VDA model below.

We adopt the following dependence of $\tau_{\rm a, \it i}$ on the mass density
and temperature of $i$th gas particle in the VDA:
\begin{equation}
\tau_{\rm a, \it i} = \tau_{\rm 0}  (\frac{ \rho_{\rm g,0} }{ \rho_{\rm g, \it i} })
(\frac{ T_{\rm g,0} } { T_{\rm g, \it i} })^{0.5},
\end{equation}
where $\rho_{\rm g, \it i}$ and $T_{\rm g, \it i}$ are the gas density and temperature
of  $i$-th  gas particle, respectively,
$\rho_{\rm a,0}$ (typical ISM density at the solar neighborhood)
and $T_{\rm g,0}$ (temperature of cold gas) are set to be 1 atom cm$^{-3}$ and 20K,
respectively, and $\tau_{\rm 0}$ is a reference dust accretion timescale
at $\rho_{\rm g,0}$ and $T_{\rm g, 0}$.
As discussed in B15,
galaxy-scale simulations like the present ones can not resolve the atomic-scale
physics of dust growth (much smaller than the possible resolution of
the present simulations). Therefore,  the above $\tau_{\rm a, \it i}$
for a gas particle should be regarded as
the dust growth timescale averaged over the smoothing length of the particle
($10-100$ pc). As shown in previous one-zone models (e.g., D98) and
our previous simulations (B13a), the dust accretion
time-scale should be an order
of $10^8$ yr so that the observed dust abundances and $f_{\rm H_2}$ can be
reproduced. We therefore adopt $\tau_{\rm a, 0} = 2 \times 10^8$ yr for all models
of the present study.

\begin{figure}
\psfig{file=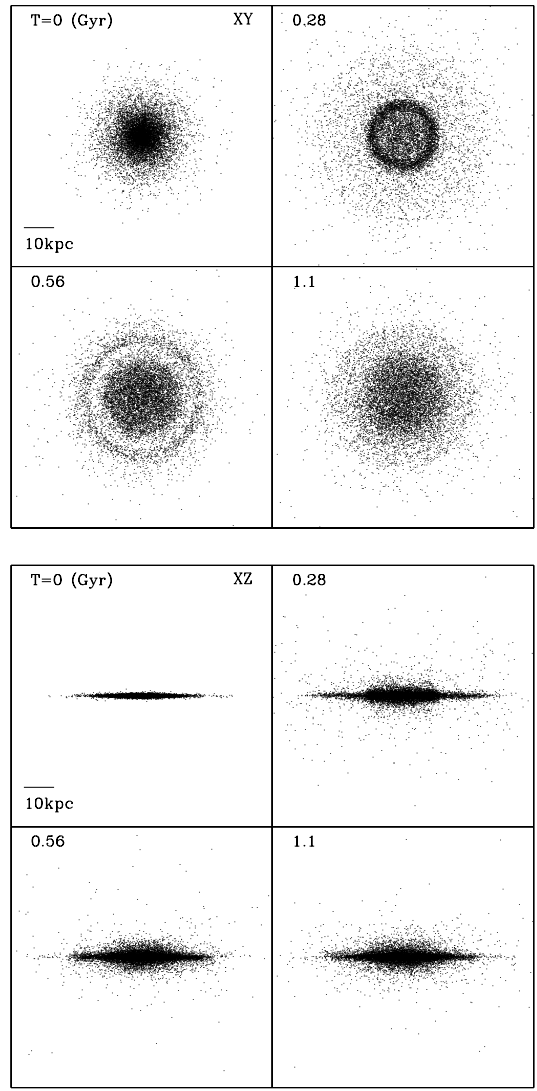,width=8.0cm}
\caption{
The time evolution of the mass distribution of silicate  projected onto the
$x$-$y$ plane (upper four) and onto the $x$-$z$ plane (lower four)
for the fiducial MW-type disk model with stellar radiation pressure.
Only one every ten particles is plotted to reduce the file size of this 
figure (yet the detail of the dust distribution can be clearly seen).
The time $T$  in the upper left corner of each panel indicates time that
has elapsed since the simulation started.
}
\label{Figure. 1}
\end{figure}

\begin{figure}
\psfig{file=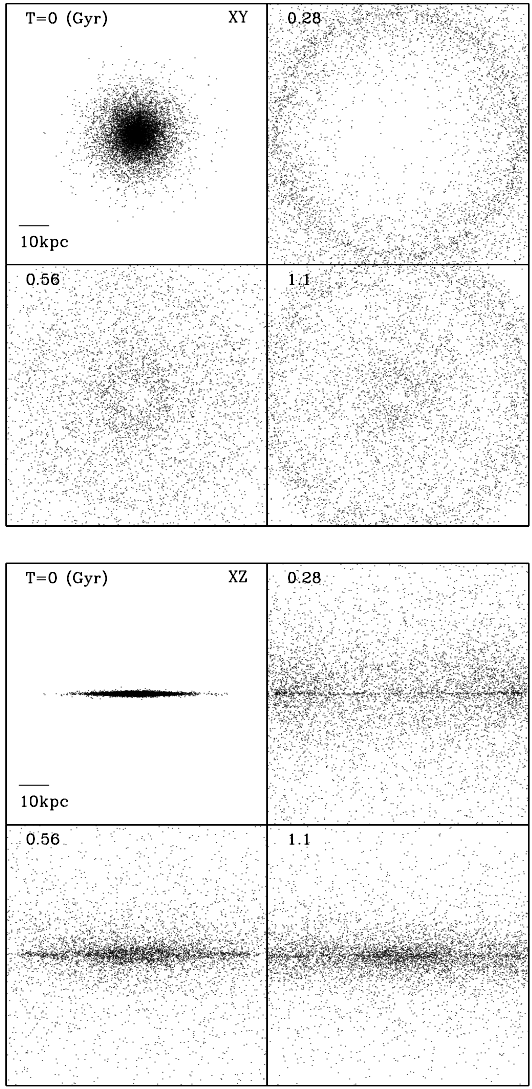,width=8.0cm}
\caption{
The same as Fig. 1 but for the distribution of graphite.
The more extended distribution of graphite in this figure is due to the combination
of the adopted small dust size and $A_{\rm t}=0$ mag. The models with
larger $A_{\rm t}$ show less extended distributions of dust.
}
\label{Figure. 2}
\end{figure}

\subsection{${\rm H_2}$ formation and dissociation}

The ${\rm H_2}$ formation and dissociation processes can be regulated
by dust evolution, because the major formation sites of ${\rm H_2}$ 
is on dust grains. The  model for these processes
 adopted in the present study
is exactly the same as those used in B13a: ${\rm H_2}$ formation
on dust grains and ${\rm H}_2$ dissociation by FUV radiation
are both self-consistently included in chemodynamical simulations.
Recent theoretical studies on  ${\rm H_2}$ formation
on dust grains have revealed the important roles of dust temperature 
and its random fluctuation in ${\rm H_2}$ formation on dust grains
(e.g. Bron et al. 2014). 
We however do not include these, mainly because dust temperature 
and its fluctuation  for each dust particle can not be reliably derived
in the present simulation without  a realistic model for dust
temperature estimation.

The temperature ($T_{\rm g}$),
hydrogen density ($\rho_{\rm H}$),  dust-to-gas ratio ($D$)
of a gas particle and the strength of the
FUV radiation field ($\chi$) around the gas particle
are calculated at each time step so that the fraction of molecular
hydrogen ($f_{\rm H_2}$) for the gas particle can be derived based on
the ${\rm H_2}$ formation/destruction equilibrium conditions.
Thus the ${\rm H_2}$ fraction for $i$-th gas  particle ($f_{\rm H_2, \it i}$)
is given as;
\begin{equation}
f_{\rm H_2, \it i}=F(T_{\rm g, \it i}, \rho_{\rm H, \it i}, D_i,  \chi_i),
\end{equation}
where $F$ means a function for $f_{\rm H_2, \it i}$ determination.
The standard value ($3 \times 10^{-17}$ cm$^3$ s$^{-1}$) for 
the total volume ${\rm H_2}$ formation rate is adopted for all models
of the present study.

Since the detail of the derivation methods of $\chi_i$ and $f_{\rm H_2, \it i}$
(thus $F$)  are given
in B13a and B15, we here briefly describe the methods.
The SEDs of stellar particles around each $i$-th gas particles
(thus ISRF) are first estimated from ages and metallicities of the stars by using stellar population
synthesis codes for a given IMF (e.g., Bruzual \& Charlot 2003).
Then the strength of the FUV-part of the ISRF
is estimated from the SEDs so that $\chi_i$ can be derived for the $i$-th gas 
particle.
Based on $\chi_i$, $D_i$, and  $\rho_{\rm H, \it i}$ of the gas particle,
we can derive $f_{\rm H_2, \it i}$ (See Fig. 1 in B13a).
Thus each gas particle has $f_{\rm H_2, \it i}$, metallicity ([Fe/H]),
and gas density, all of which are used for estimating the IMF slopes
for the particle (when it is converted into a new star).
The ages of stars ($t_{\rm 0}$) are assumed to be free parameters
ranging from 0.1 Gyr and 10 Gyr in the present study.

\begin{table}
\centering
\begin{minipage}{80mm}
\caption{Description of the fiducial model.}
\begin{tabular}{ll}
Disk model & MW-type \\
Dust type & silicate (S) or graphite (G)  \\
RP  &  Yes  \\
PEH  &  No \\
GD  &  No \\
CR  &  No \\
DR  &  No  \\
Dust growth  &  Yes \\
Dust destruction   &  No \\
$R_{\rm grow}$ & $0.5 \epsilon_{\rm g}$ \\
$A_{\rm t}$   &  0 mag \\
$t_{\rm 0}$   &  5 Gyr \\
$N_{\rm dust}$   &  $10^5$ \\
$\rho_{\rm dust,S}$   &  $3.0$ g cm$^{-3}$ \\
$R_{\rm dust,S}$   &  $0.1$  $\mu$m \\
$\rho_{\rm dust,G}$   &  $2.3$ g cm$^{-3}$ \\
$R_{\rm dust,G}$   &  $0.05$  $\mu$m \\
IMF   &  Salpeter ($\alpha_{\rm IMF}=-2.35$)  \\
SSP & Bruzual \& Charlot (2003)
\end{tabular}
\end{minipage}
\end{table}

\begin{table}
\centering
\begin{minipage}{80mm}
\caption{The adopted values of four key parameters.}
\begin{tabular}{ll}
Parameter & Adopted values \\
Dust type &  silicate, graphite \\
$A_{\rm t}$  &  0, 0.3, 1, 2 (mag) \\
$t_{\rm 0}$  &  1, 3, 5,  10 (Gyr) \\
$Q_{\rm rp}^{\ast}$   &   0, 0.5, 1, 2 \\
\end{tabular}
\end{minipage}
\end{table}

\subsection{Photo-electric heating by dust}

Heating of ISM through the photo-electric (`PEH') ejections of electrons from
dust grains interacting with ISRF has been discussed by many theoretical
works (e.g., Watson 1972;  Draine 1978; Bakes \& Tielens 1994, BT94). 
These works clearly demonstrated that the photo-electric heating is important
in the thermal history of ISM and the level of this importance depends
on the detailed physical conditions of ISM, such as the strength of ISRF
and dust size distributions (e.g., BT94). 
We here include this important photo-electric heating in our galaxy-scale
simulations in such a way that the net heating rate is estimated for each gas
particle according to the physical properties of the gas particle and those of dust
particles around the gas particle. Since our estimation of the photo-electric
heating rate is estimated directly from gas and dust properties,
our model is much better than those adopted in other galaxy-scale simulations
with photo-electric heating (e.g., Tasker 2011) in which the evolution of ISRF
and dust abundances is not considered.

We adopt the analytic formula for photo-electric heating rate ($n\Gamma_{\rm pe}$)
proposed by BT94 as follows:
\begin{equation}
n\Gamma_{\rm pe} = 1.0 \times 10^{-24} \epsilon n G_0
\; {\rm ergs}  \; {\rm cm^{-3}} \; {\rm  s^{-1} },
\end{equation}
where $n$ is the number density of gas, $\epsilon$ is the heating efficiency,
and $G_0$ is the intensity of the incident far-UV field in units of 
Habing interstellar radiation field. BT94 showed  that (i) $\epsilon$ depends mainly on
$G_0$, $T_{\rm g}$, and $n_{\rm e}$, where $n_{\rm e}$ is the electron number
density, in ISM (e.g., equation 43 in BT94) and (ii) 
$\epsilon$ ranges from 0.001 to 0.05 within their models.
Although our simulations can output $G_0$, $n$, and $T_{\rm g}$, $n_{\rm e}$ is
not the direct output of our simulations.
We therefore assume that $\epsilon$ is a parameter for a model in the present
study. We mainly show the results of the models with $\epsilon=0.003$
in the present study, and our forthcoming papers will discuss how $\epsilon$
can control thermal evolution of ISM in galaxies.

In estimating $G_0$ for each gas particle, we need to consider that
not all of the FUV flux of a star around the gas particle can be used
for photo-electric heating owing to dust extinction. 
The flux at a wavelength $\lambda$
for $i$th star around $j$th gas particle in a screen
model can be given as follows
\begin{equation}
f_{\lambda, i}=f_{\lambda, 0, i} e^{-\tau_{\lambda, j} r_{i,j}/h_j},
\end{equation}
where $f_{\lambda, 0, i}$ is the original flux of the star,
$\tau_{\lambda,j}$ is the optical depth
(i.e. $\tau_{\lambda} \approx 0.921 A_{\lambda}$), $r_{i,j}$ is the distance of
the gas and the star, and $h_j$ is the SPH smoothing length
of the gas particle. Therefore, the $G_0$ factor for photo-electric heating
of $j$th gas particle from $i$th stellar particle ($G_{0,i,j}$) is
\begin{equation}
G_{0,i,j}=F_{\rm ext, \it i, j} g_{0,i,j}, 
\end{equation}
where $g_{0,i,j}$ is $G_0$ estimated by assuming no dust extinction
and $F_{\rm ext, \it i,j}$ is the fraction of light absorbed
by dust and given as follows:
\begin{equation}
F_{\rm ext, \it i, j} = \int_0^{1} e^{-\tau_{\it j}  r}  dr,
\end{equation}
where $\tau_j$ is the FUV optical depth
and a fixed value  is used.
Since the adopted SPH kernel is 0 at $h_j$, the above integration range
needs to be from 0 to 1.
In deriving $\tau_{\it j}$, we first estimate the optical dust
extinction ($A_{\rm V, \it j}$) based on the gas column density
($N_{\rm H, \it j}$)  and 
the dust-to-gas-ratio ($D_j$). Then we estimate $A_{\rm FUV, \it j}$
from $A_{\rm V, \it j}$ by using the Calzett's extinction law
(i.e., the equation (4) in Calzetti et al. 2000).
In estimating $A_{\rm V, \it j}$,
we  use the following observational results around
the solar neighborhood (Predehl \& Schmitt 1995):
\begin{equation}
\frac{ N_{\rm H} }{ A_{\rm V} }=
1.8 \times 10^{21} \;  {\rm atoms \; cm^{-2} \; mag^{-1}  }.
\end{equation}
Accordingly, $A_{\rm V, \it j}$ is linearly proportional to
$N_{\rm H, \it j}$ and $D_{\it j}$. The ratio of $A_{\rm FUV}$ to $A_{\rm V}$
is estimated to be 2.56 and this fixed value is used for all gas particles
to derive $A_{\rm FUV}$ (thus $\tau_{\rm FUV}$).
The local  column density $N_{\rm H, \it j}$ is estimated as 
$\rho_j{\rm (H)} h_j$, where $\rho_j{\rm (H)}$ is the 3D hydrogen density
and $h_j$ is the smoothing length for the SPH gas particle.
Thus $G_0$ can be estimated from physical properties of gas and dust in
a self-consistent manner.

\begin{figure}
\psfig{file=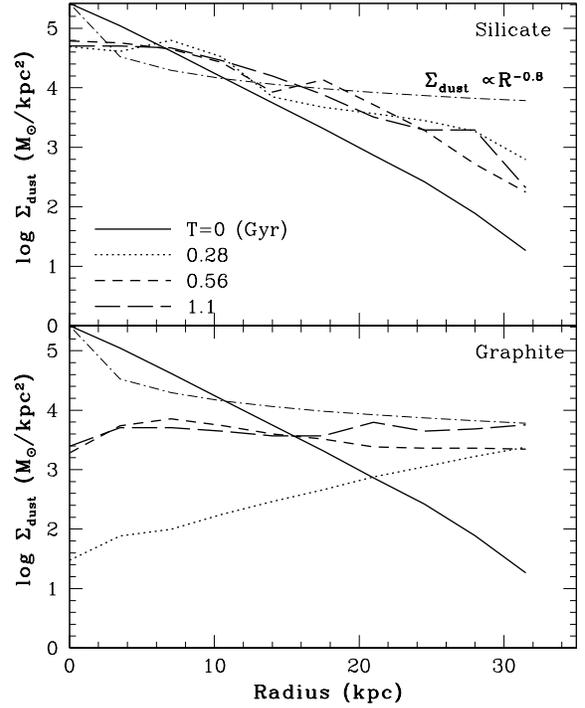,width=8.0cm}
\caption{
The projected radial mass density profiles of silicate (upper) and
graphite (lower) at four selected time steps in the fiducial MW-type disk
model. The radius ($R$) in this figure is the distance from the center of
galaxy in the $x$-$y$ plane ($R=\sqrt{x^2+y^2}$).
The solid, dotted, short-dashed, and long-dashed lines indicate
the dust distributions at $T=0$, 0.26, 0.56, and 1.1 Gyr, respectively.
For comparison, the observed rather flat distribution of dust around
galaxies (M10) is shown by
a dot-dashed line with an arbitrary $\Sigma_{\rm dust}$ at $R=0$.
}
\label{Figure. 3}
\end{figure}

\subsection{Cosmic-ray heating}

Although gas heading by cosmic-ray (`CR') would not be  so efficient as photo-electric
heating in typical  ISM of star-forming galaxies 
($n \sim 1$ atom cm$^{-3}$, SFR$\sim 1$ ${\rm M}_{\odot}$ yr$^{-1}$),
it can be important in thermal histories in some galaxy environments
(e.g., central starburst regions). We therefore include this cosmic-ray heating
by using the following analytic formula described in Tielens (2005);
\begin{equation}
n\Gamma_{\rm CR} =  3.0 \times 10^{-27} n
( \frac{ \zeta_{\rm CR} }{2 \times 10^{-16} })  
\; {\rm ergs}  \; {\rm cm^{-3} } \; {\rm  s^{-1} },
\end{equation}
where $\zeta_{\rm CR}$ is total  cosmic ionization rate.
If cosmic ray originates from SNe, then it is reasonable
for us to assume that the total flux of comic ray in a galaxy  is proportional
to the SFR of the galaxy. We therefore assume that
$\zeta_{\rm CR}$ is $2 \times 10^{-16} (\frac{ {\rm SFR} }
{ {\rm 1.3 {\rm M}_{\odot} yr^{-1} } })$ s$^{-1}$,
where the reference SFR of $1.3 {\rm M}_{\odot}$ yr$^{-1}$ corresponds
to the present SFR of the Galaxy (e.g., Draine 2009).
We admit that there is a great uncertainty in the scaling relation between
$\zeta_{\rm CR}$ and SFRs of galaxies. Accordingly, it would be better for
the present study to discuss the roles of dust in the evolution of ISM
by {\it not} including cosmic-ray heating.

\subsection{Gas-dust  heating}

In principle, we can estimate separately the temperature of gas
($T_{\rm g}$)  and that of dust 
($T_{\rm d}$) in the present study with the new live dust particle method. If 
$T_{\rm d}$ is higher than $T_{\rm g}$, then ISM can be heated up. 
Given that the observed mean $T_{\rm d}$ is around $20-30$K (Dunne et al. 2001)
and thus not so different,
this dust-gas heating would  be important only for a small amount of gas
in galaxies.  We adopt the following formula (Tielens 2005) for the estimation
of dust-gas cooling rate:
\begin{equation}
n\Gamma_{\rm gd} =  1.0 \times 10^{-33} n
T_{\rm g}^{0.5} (T_{\rm d}-T_{\rm g})
\; {\rm ergs}  \; {\rm cm^{-3} } \; {\rm  s^{-1} }.
\end{equation}
In order to derive $T_{\rm d}$, we use the observed scaling relation
between $T_{\rm d}$ and $L_{\rm FIR}$ (total FIR luminosity) by
Amblard et al. (2010) as follows:
\begin{equation}
T_{\rm d} \; ({\rm K}) = -20.5 + 4.4 \log (L_{\rm FIR}/L_{\odot}).
\end{equation}
We estimate  $L_{\rm FIR}$  from the observed relation between
SFR and $L_{\rm FIR}$ (Kennicutt 1998) as follows:
\begin{equation}
\frac{ {\rm SFR} }{ {\rm M}_{\odot} {\rm yr^{-1}} } 
= \frac{ L_{\rm FIR} }{ 5.8 \times 10^9 {\rm L}_{\odot} }.
\end{equation}

In the above estimation of $\Gamma_{\rm gd}$, the dust temperature
of a galaxy is estimated by using the galactic SFR and 
$T_{\rm d}$ is used for all gaseous particles. 
This method  is clearly over-simplified, though such a simple model
could possibly grasp some essential ingredients of dust-gas heating
in ISM evolution. We therefore do not include the adopted dust-gas
heating in most of the present simulations. We just briefly discuss
how the possible effects of dust-gas heating on ISM are for a
number of models. As Bekki \& Shioya (2000) has already shown,
$T_{\rm d}$ can be  estimated for each particle by using radiation-transfer
models for arbitrary geometry of a galaxy modeled by Nbody + gas dynamical
simulations. In our forthcoming papers, we need to estimate $T_{\rm d}$
for each dust particle so that we can discuss the roles of dust-gas
heating in ISM evolution in a more convincing manner.

\begin{figure*}
\psfig{file=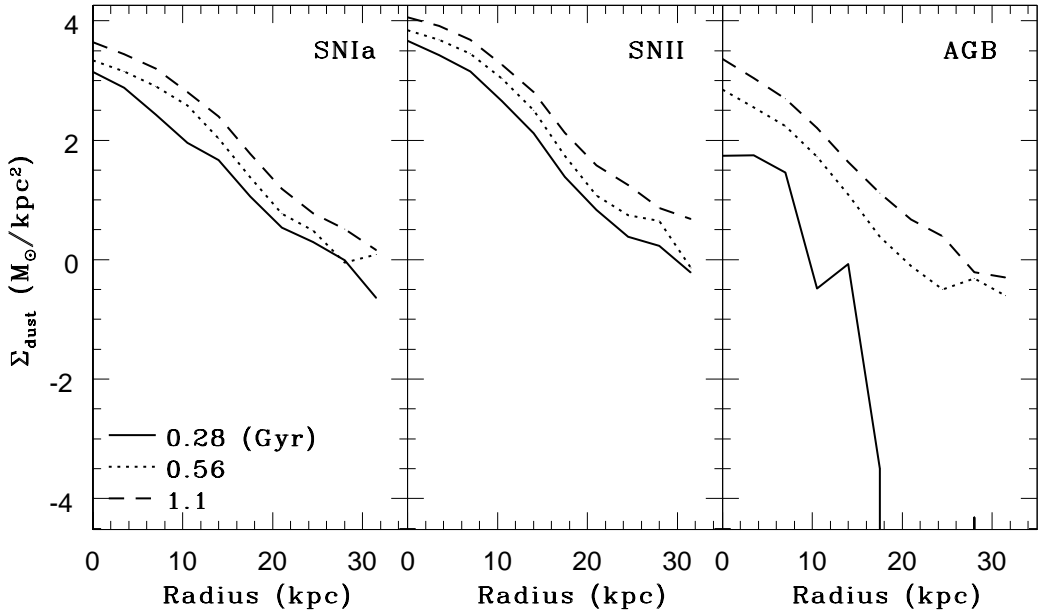,width=14.0cm}
\caption{
The projected radial mass density profiles of newly produced dust from
SNIa (left), SNII (middle), and AGB stars (right) in the fiducial MW-type
disk model. The solid, dotted, and short-dashed lines indicate
the dust profiles at $T$=0.28, 0.56, and 1.1 Gyr, respectively.
}
\label{Figure. 4}
\end{figure*}

\begin{figure*}
\psfig{file=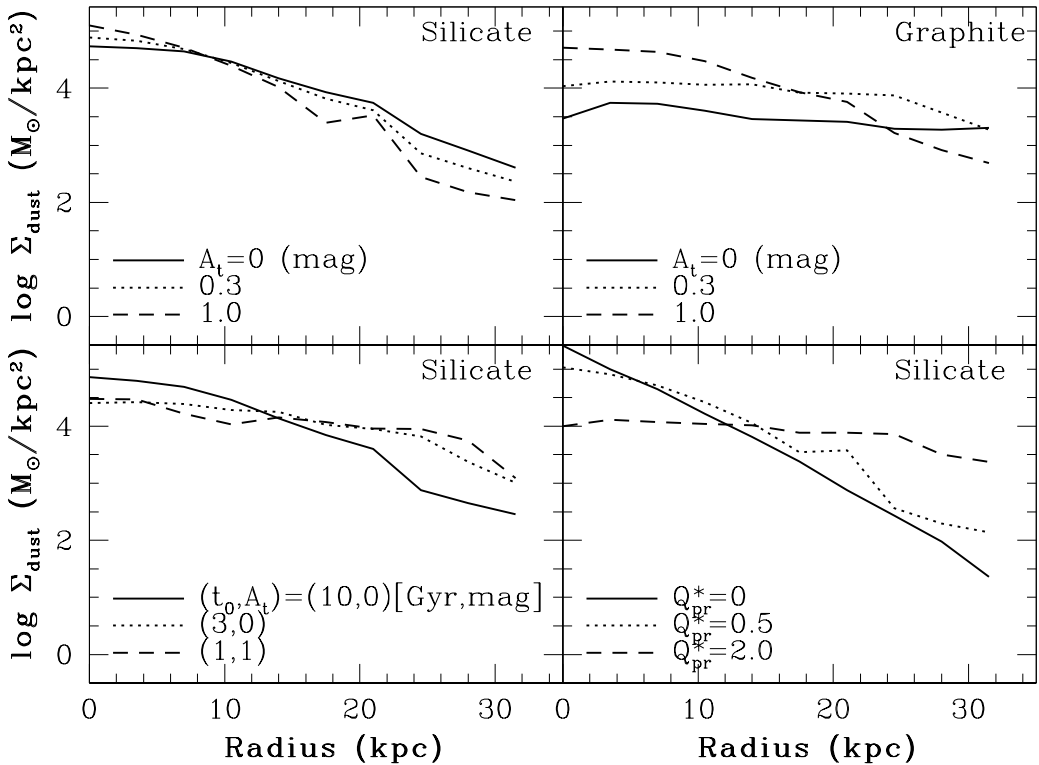,width=14.0cm}
\caption{
The final projected radial mass profiles of dust in different 12 MW-type
disk models with stellar radiation pressure. The upper left and right panels
shows the profiles of silicate and graphite,
respectively, for $A_{\rm t}=0$,
(solid), 0.3 (dotted), and 1.0 mag (short-dashed).
The lower left panel shows the profiles of silicate
in the models with  
$(t_0,A_{\rm t}$=(10,0) (solid)
=(3,0) (dotted)
(1,1) [Gyr, mag] (solid)
(solid), 0.3 (dotted), and 1.0 mag (short-dashed).
The solid, dotted, and short-dashed lines in the lower right
panel indicate $Q_{\rm pr}^{\ast}=0$,
0.5, and 2.0, respectively.
The model with  $Q_{\rm pr}^{\ast}=0$ corresponding to 'no radiation pressure'.
}
\label{Figure. 5}
\end{figure*}

\begin{figure*}
\psfig{file=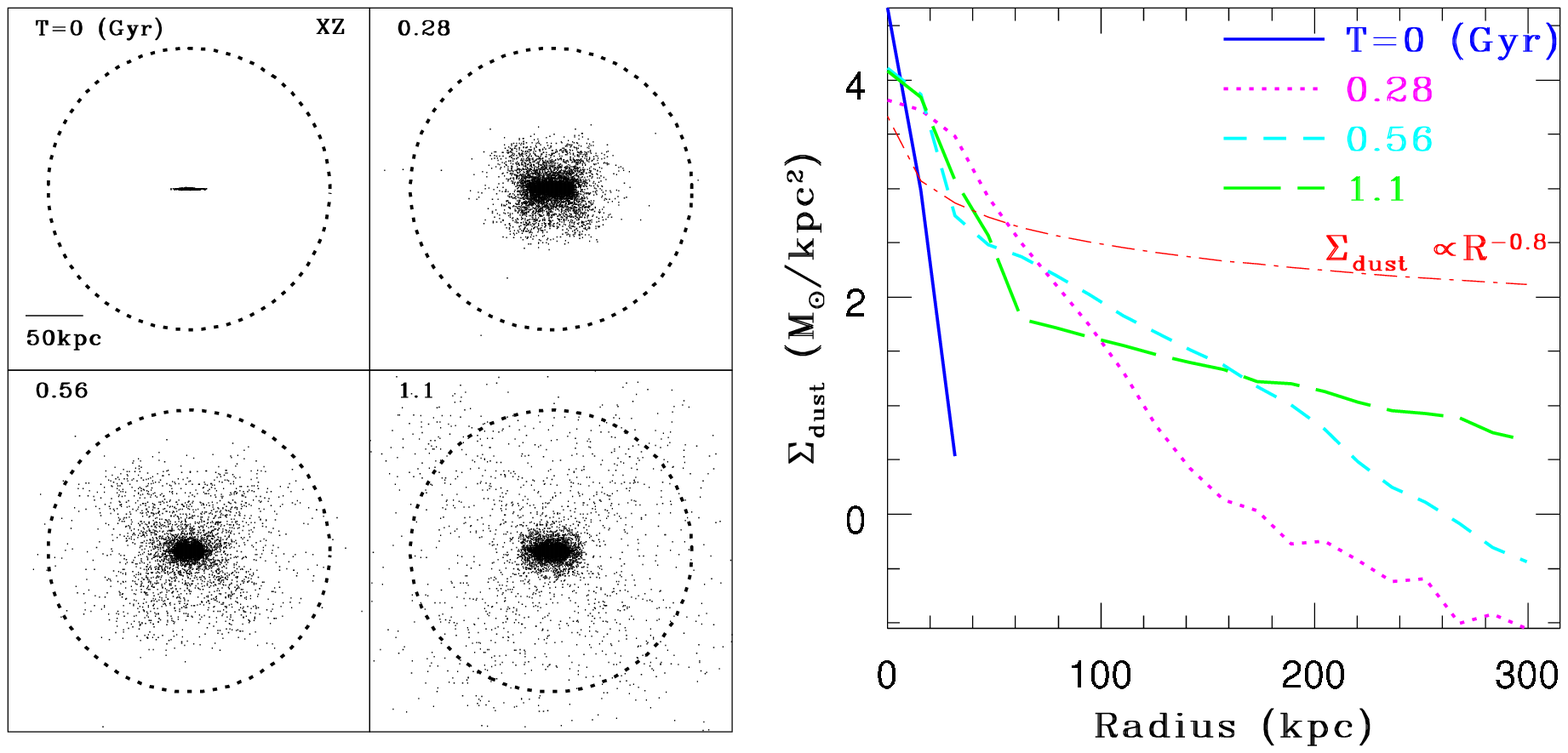,width=14.0cm}
\caption{
The time evolution of graphite distributions  projected onto the $x$-$z$ plane
in the MW-type disk model with radiation pressure and 
$A_{\rm t}=0.3$ mag (left) and the projected radial density profiles
of graphite for the selected four time steps (right). The virial radius
of the dark matter halo is shown by a thick dotted line in the left panel.
The blue solid,
magenta dotted, cyan short-dashed, and green long-dashed lines indicate
$T=0$, 0.28, 0.56, and 1.1 Gyr, respectively in the right panel. For comparison,
the observed profile of dust around galaxies by M10 is shown by a red dot-dashed
line with an arbitrary $\Sigma_{\rm dust}$ at $R=0$.
}
\label{Figure. 6}
\end{figure*}

\subsection{Gaseous drag effects on dust}

Theis \& Orlova (2004) first investigated the roles of dust in dynamical
evolution of ISM in the central regions of galaxies and thereby found that
the central gas disks can become dynamically unstable if the mass fractions
of dust in the disks exceed 2\%.
These results imply that gas-dust coupling by a drag force in ISM can be
an important aspect of ISM evolution in galaxies. However, the frictional (drag)
timescale ($\tau_{\rm drag}$,
which corresponds to the timescale for momentum transfer between gas 
and dust through gas-dust collision) is rather short (an order of $10^3-10^4$ yr)
for ISM with typical dust sizes and masses.
The maximum time-step width, $\Delta t_{\rm max}$ ($ \sim 10^6$ yr),
in the present simulations
is significantly longer than $\tau_{\rm drag}$.
Therefore,
it would be reasonable for us to assume that 
even if gas and dust initially have different velocities, 
they come to have 
identical velocities within $\Delta t_{\rm max}$. 

However,
$\tau_{\rm drag}$ can be longer than $\Delta t_{\rm max}$
in some local regions during the evolution of ISM. We accordingly
investigate how the gaseous drag can influence the evolution of dust 
in galaxies by adopting the model below.
The frictional drag force ($F_{\rm drag}$) between gas and dust 
is described as follows:
\begin{equation}
{\bf F}_{\rm drag}=-C_{\rm drag}\eta_{\rm dust}( {\bf v}_{\rm d} - {\bf v}_{\rm g} ),
\end{equation}
where ${\bf v}_{\rm d}$ and ${\bf v}_{\rm g}$ are the velocity vectors
of dust and gas, respectively,
and $\eta_{\rm dust}$ is
$m_{\rm d}/m_{\rm dust}$ (where $m_{\rm d}$ is the dust particle mass and $m_{\rm dust}$
is the mass of a  dust grain): this $\eta_{\rm dust}$ needs to be considered,
because this $F_{\rm drag}$ is for a dust partilce (not for a dust grain).
The coefficient
$C_{\rm drag}$ corresponding to $\tau_{\rm drag}^{-1}$
(Noh et al. 1991; Theis \& Orlova 2004) 
is given as
\begin{equation}
C_{\rm drag}= \frac{ \sigma_{\rm c} \rho_{\rm g} v_{\rm th} }
{ m_{\rm dust} },
\end{equation}
where $\sigma_{\rm c}$,
$\rho_{\rm g}$, $v_{\rm th}$, and $m_{\rm dust}$
are the cross section of a dust grain
($\sigma_{\rm c}=\pi R_{\rm dust}^2$, where $R_{\rm dust}$ is the dust size),
the mass density
of the gas, the thermal velocity of the gas,
and the total mass of the grain. It should be stressed here
 that this $m_{\rm dust}$
is not the mass of a dust particle in a simulation.

In implementing the above drag effects for each dust particle in a simulation,
we need to consider that $F_{\rm drag}$ for $i$th dust particle 
($F_{\rm drag, \it i}$) is the force from all gas particles
around the dust particle, as follows:
\begin{equation}
{\bf F}_{\rm drag, \it i} = 
{\bf f}_{\rm drag, \it i, j},
\end{equation}
where ${\bf f}_{\rm drag, \it i, j}$ is the drag force between $i$th dust 
particle and $j$th
gas particle that is the nearest to the dust particle.
This $f_{\rm drag, \it i, j}$ is 
calculated from the velocities of the particles and 
$C_{\rm drag}$. The back-reaction of the drag force 
needs to be considered for  $j$th gas particle  because of momentum conservation
as follows:
\begin{equation}
{\bf F}_{\rm drag, \it j} = -m_{\rm d, \it i} {\bf f}_{\rm drag, \it i, j}
/m_{\rm g, \it j}.
\end{equation}
These drag terms are added to the equations of motion at each time step.

\begin{figure*}
\psfig{file=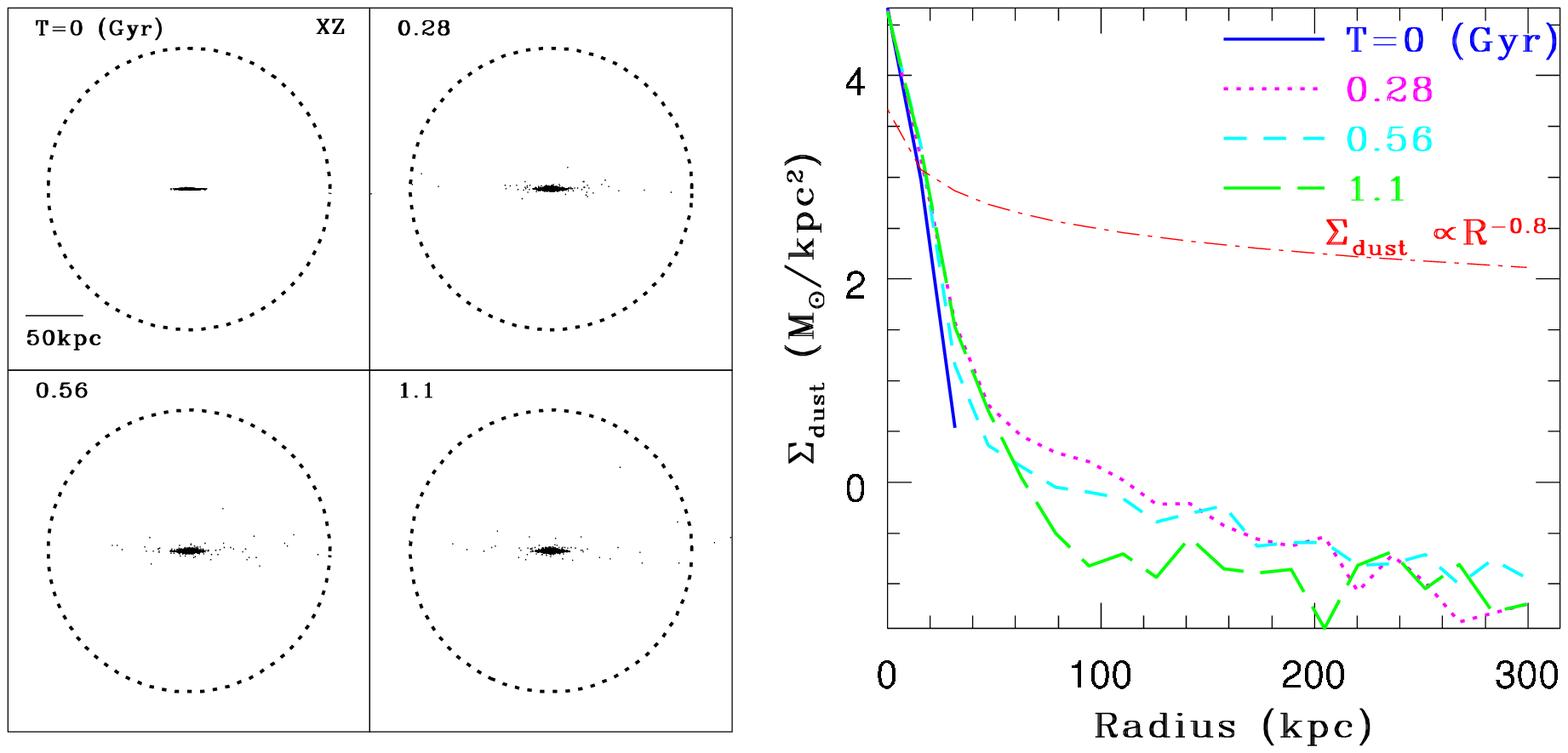,width=14.0cm}
\caption{
The same as Fig. 6 but for the silicate distribution
in the MW-type disk with radiation pressure
and heavy dust extinction ($A_{\rm t}=1$ mag). It should be stressed here that
dust can not be so extended in the vertical direction owing to the much weaker
stellar radiation field in this model.
}
\label{Figure. 7}
\end{figure*}

\subsection{Radiation pressure of stars on dust}

Barsella et al. (1989) and F91 investigated how radiation
pressure of stars on dust grains can influence their orbital evolution
around their host galaxy by using some idealized models
in which both the gravitational potential
and the light distribution of the galaxy are fixed during dust evolution.
Here we improve their model by calculating the light
and mass  distributions of a galaxy
at every time step so that we can better estimate the 3D radiation field
of the galaxy. For each $i$th dust particle, 
$F_{\rm rad, \it i}$ is the sum of the force due to radiation pressure
from all stars as follows (e.g., Barsella et al. 1989):
\begin{equation}
{\bf F}_{\rm rad, \it i}= 
\frac{ \pi \eta_{\rm dust} R_{\rm dust}^2 Q_{\rm pr}^{\ast} }{c}
\sum_{j=1}^{N_{\rm s}} L_{\rm s, \it j}
\frac{ {\bf  x}_i  - {\bf   x}_j}
{ 4 \pi |{\bf  x}_i  - {\bf  x}_j|^3 },
\end{equation}
where  $Q_{\rm pr}^{\ast}$ is the frequency-averaged radiation pressure coefficient
for a grain, $c$ is the speed of light,
$N_{\rm s}$ is the total number of stellar particles,
$L_{\rm j}$ is the total luminosity of $j$th stellar particle,
${\bf  x}_i$ is the 3D position vector of $i$th dust particle,
and ${\bf  x}_j$ is the 3D position vector of $j$th stellar  particle.
Since this is $F_{\rm rad}$ for a dust particle, the $\eta_{\rm dust}$ factor
needs to be considered.

In order to estimate $L_{\rm s, \it j}$ for $j$th stellar
particle with a metallicity and an age, we use the stellar population
synthesis code by Bruzual \& Charlot 2003) for the adopted IMF.
By considering dust extinction,
we can estimate $L_{\rm s, \it j}$ as follows:
\begin{equation}
L_{\rm s, \it j}=m_{\rm s, \it j}\Upsilon_j^{-1} e^{-0.921A_j},
\end{equation}
where $\Upsilon_j$ is the mass-to-light-ratio derived from the stellar population
synthesis code for $j$th stellar particle and $A_j$ is the dust extinction
for the particle.
If we estimate $A_j$ in a fully self-consistent manner,
we need to do so by investigating the column densities and dust abundances
of all gas particles that the light of $j$th stellar particle 
can pass through while it is traveling to $i$th dust particle.
This means that the required time for the  calculation of
$F_{\rm rad}$ for just one
dust particle is proportional to $N_{\rm s} N_{\rm g}$:
thus the calculations for all dust particles at each time step
is proportional to $N_{\rm d}N_{\rm s}N_{\rm g}$ (virtually $\propto N^3$).
This calculation is very numerically costly and thus infeasible
unless a fast method for this calculation is developed.
We therefore investigate the models with a fixed (total) dust extinction ($A_{\rm t}$)
in which $A_{\rm t}$ is 0, 0.3, 1. and 2 mag. By comparing the model with different
$A_{\rm t}$, we can better understand how dust extinction can be important
for the evolution of dust being influenced by radiation pressure of stars
in galaxies.

Following F91, we investigate mainly two dust species:
interstellar silicate with $R_{\rm dust} \sim 0.1$ $\mu m$ and 
$\rho_{\rm dust} = 3.0$ g cm$^{-3}$ and graphite  with
$R_{\rm dust} \sim 0.05$ $\mu m$ and $\rho_{\rm dust} = 2.3$ g cm$^{-3}$.
In order to demonstrate the roles of radiation pressure of stars in the
evolution of dust more clearly,
interstellar dust is assumed to be
either of the above silicate or graphite 
in the present simulations. 
Given that interstellar dust consists of different grains with different
compositions and sizes,  
this is obviously an over-simplification. 
Although we admit this,  we think that we need to start this investigation
with a somewhat idealized model in order to grasp some essential ingredients
of stellar radiation pressure on dust dynamics. We therefore 
try to understand the influences of stellar radiation pressure
on dust evolution by using a simpler model  in the present study,
and  will construct a more realistic and complicated model for the influences
on dust with different masses and sizes
in our future papers.

As a galaxy evolves, $Q_{\rm pr}^{\ast}$ 
for a given grain also changes according to the 
time evolution of its SED. It is therefore self-consistent
and ideal  for the present study to 
estimate $Q_{\rm pr}^{\ast}$ at each time step by considering the SED evolution of
a galaxy. However, this self-consistent estimation of time-varying $Q_{\rm pr}^{\ast}$
could make the present simulated code very complicated, because the 
derivation of the dust-corrected
SED of a simulated  galaxy is not a simple task (e.g., Bekki \& Shioya 2000).
We therefore adopt an idealized model in which  $Q_{\rm pr}^{\ast}$ for a dust grain
is constant all over the time in a simulation. 
We mainly investigate 
the evolution of  disk galaxies with a smaller gas fraction
only for $\sim 1$ Gyr for most models. Therefore,
this assumption might be justified, because the SED does not change 
drastically owing to the lack of major bursts of star formation.

Barsella et al. (1989) derived $Q_{\rm pr}^{\ast}$ by assuming a single
black body temperature and listed the values for different
temperatures in their Table 1. F91 considered the SEDs of galaxies with
Sb and Sc Hubble types and estimated $Q_{\rm pr}^{\ast}$ and showed
the dependence of $Q_{\rm pr}^{\ast}$ on dust grain sizes in their Fig. 1.
The estimated $Q_{\rm pr}^{\ast}$ of silicate  with
$R_{\rm dust}=0.1$ $\mu$m is $\sim 1$ 
and that of graphite grains with $R_{\rm dust}=0.05$ $\mu$m is $\sim 1$
too for late-type disk galaxies. Although we use these values
as a reference,
we also  investigate how the present results depend on $Q_{\rm pr}^{\ast}$ by using
the models with different $Q_{\rm pr}^{\ast}$.

\subsection{Dust-related processes not included in this work}

It is important for the present study to clearly indicate  the dust-related
physical processes that are not included in the new live dust particle
model (Table 1 briefly summarize this). The following three important
processes are not modeled in the present study. The first is the 
size distribution of dust and its evolution in ISM of galaxies.
The dust size distribution, which is a key parameter for the SEDs
of galaxies, can be significantly evolved owing to selective
destruction of smaller dust grains (e.g., Asano et al. 2014).
The second is the dust evolution due to coagulation and shattering of dust grains
in ISM, which is included in recent one-zone models (e.g., Hirashita 2012).
Given that the net formation efficiency of ${\rm H_2}$ on dust grains in
ISM of  a galaxy
depends strongly on the dust size distribution,
not including these two could underestimate or overestimate ${\rm H_2}$ 
contents of galaxies.

Third is the destruction of dust by warm/hot gaseous halos of galaxies, 
the detailed processes of which would 
depend on the mass-density and temperature of gaseous halos.
This non-inclusion of dust destruction
can result in the overestimation
of the total dust mass in the halo regions of galaxies. 
The lack of the detailed  observational information on the 3D distributions
of gas density and temperature in the Galactic halo currently prevents us from
discussing the level of this possible overestimation of the dust mass.
Other possibly important factors, such as magnetic fields and turbulence of ISM,
are not included either in this work, because they could play a relatively minor
role in comparison with other factors discussed in the present work.

The size distributions of dust can be influenced by dust formation, destruction, growth,
coagulation, and shattering processes, and thus SEDs can be influenced by  these processes
too. The radiation pressure 
of stars on dust grains is determined by $Q_{\rm pr}^{\ast}$
and thus by dust-corrected SEDs.  Therefore, ultimately speaking,
both (i) radiative-transfer of stellar light in dusty ISM and 
(ii) dust-related physical processes
need to be self-consistently solved at each time step in a simulation.
This means that the present code is not so fully self-consistent and our future 
simulation codes will need to incorporate above two complicated ingredients
of dusty ISM evolution for modeling ISM of galaxies in a much more self-consistent
way.

\begin{figure*}
\psfig{file=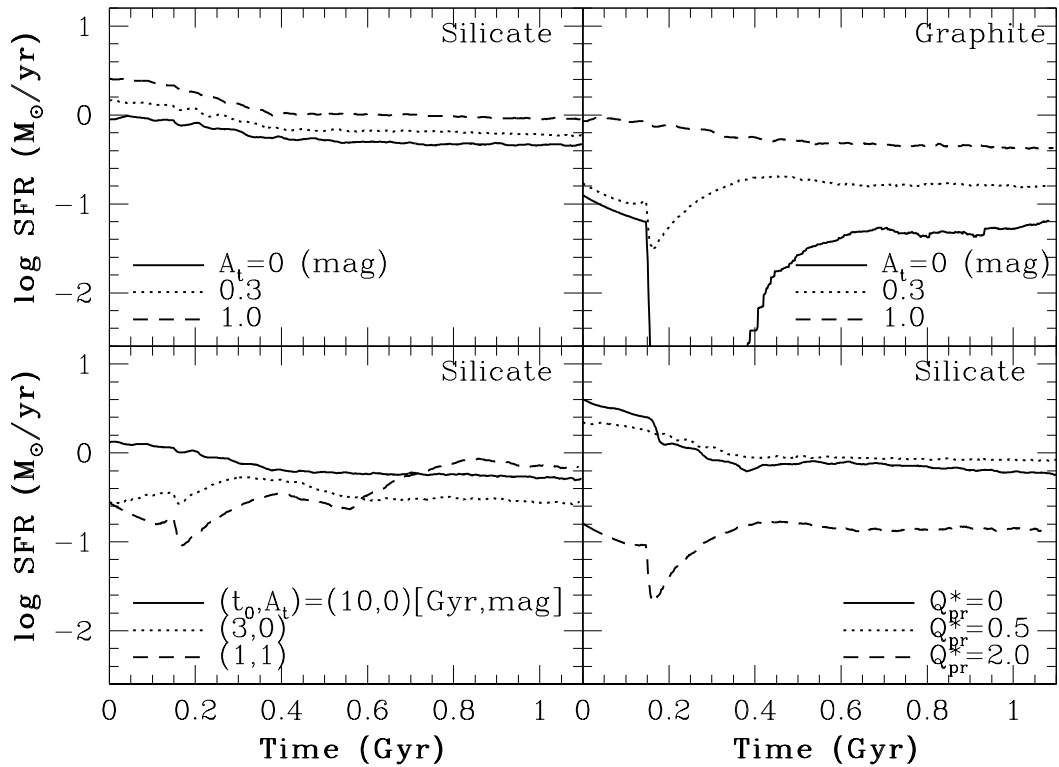,width=14.0cm}
\caption{
The time evolution of SFRs in the 12 representative models. 
The line types are exactly the same as those used in Fig. 5.
}
\label{Figure. 8}
\end{figure*}

\subsection{Parameter study}

\subsubsection{Fiducial MW-type disk galaxy model}

We mainly investigate the MW-type disk model 
with $M_{\rm h}=10^{12} {\rm M}_{\odot}$
in which spatial distributions
of gas and stars are consistent with the observed one for the Galaxy.
The adopted model parameters such as $M_{\rm s}$ and  $M_{\rm g}$
are given in Table 4. 
Since the main purpose of this paper is not to describe the dependences
of the results on the physical properties of disks, we investigate
only three other disks models.
These are  `LMC-type' (referred to as `LMC')
with $M_{\rm h}=10^{11} {\rm M}_{\odot}$,
`Dwarf-type' (`DW'),
with $M_{\rm h}=10^{10} {\rm M}_{\odot}$,
and `Sa-type' 
with $M_{\rm h}=10^{12} {\rm M}_{\odot}$
and a big bulge with $M_{\rm b}=1.2 \times 10^{11} {\rm M}_{\odot}$
(i.e., B/T=0.67 corresponding to early-type spirals).
These names  are used just for
distinguishing three disk galaxies with different $M_{\rm h}$. 
The model parameters for these disk models are also
given in Table 3.

First we show the results of the  fiducial MW-type disk model with
stellar radiation pressure on dust grains in order to discuss
the basic roles of the radiation pressure on dust evolution in \S 3.1.
We do not include dust-related  physical processes other than
stellar radiation pressure in the fiducial model, 
because we need to show the basic roles
of dust as clearly as possible.
Then we describe how the present results can
depend on other model parameters in \S 3.2.
The basic parameters and physical processes included (or not included)
in the fiducial model are given in Table 4.
The initial total numbers of particles used in the fiducial MW-type,
LMC-type,  Dwarf-type, and Sa-type disk models
are 1134000, 1100000, 1100000, and 1234000,
respectively. The total particle  numbers can increase as star formation creates
new SNe and AGB stars.

\subsubsection{Key parameters}

The main purpose of this paper is to 
understand the roles of radiation-driven dust wind
in the evolution of dust, star formation rates (SFRs), and molecular
fractions ($f_{\rm H_2}$), and chemical abundances of galaxies.
Therefore we first investigate these roles  by using
the fiducial MW models with and without radiation pressure of stars
on dust. In these models,  the effects of photo-electric heating,
dust-gas heating, gaseous drag, and cosmic-ray heating are not included
so that we can more clearly understand the roles of
radiation-driven dust wind in galaxy evolution.
The key parameters in this first study is the initial ages of stars
in the disks ($t_0$), $Q_{\rm pr}^{\ast}$,  dust-types,
and $A_{\rm t}$ (total dust extinction). Then we investigate different
disk models with and without different dust physical effects in order
to confirm whether  the roles of radiation-driven dust wind in galaxy evolution 
in these models can be clearly seen in the models.
The adopted values of the four key parameters are shown in Table 5.

\subsubsection{Photo-electric heating and gas-dust drag}

Although we have found important roles of photo-electric heating of ISM
by dust in the star formation histories
(SFHs) of galaxies, we do not discuss these extensively
in the present study, and will describe them in detail in our forthcoming
papers. This is mainly because we need to discuss this issue thoroughly
by using the results of many models and thus it is not so appropriate
for the present study to include them in this first paper. 
We have also found the effects of gas-dust drag on the evolution
of the 3D spatial distributions of gas and dust in galaxies. 
We however discuss these in our forthcoming paper for the same reason
as above.

\begin{figure*}
\psfig{file=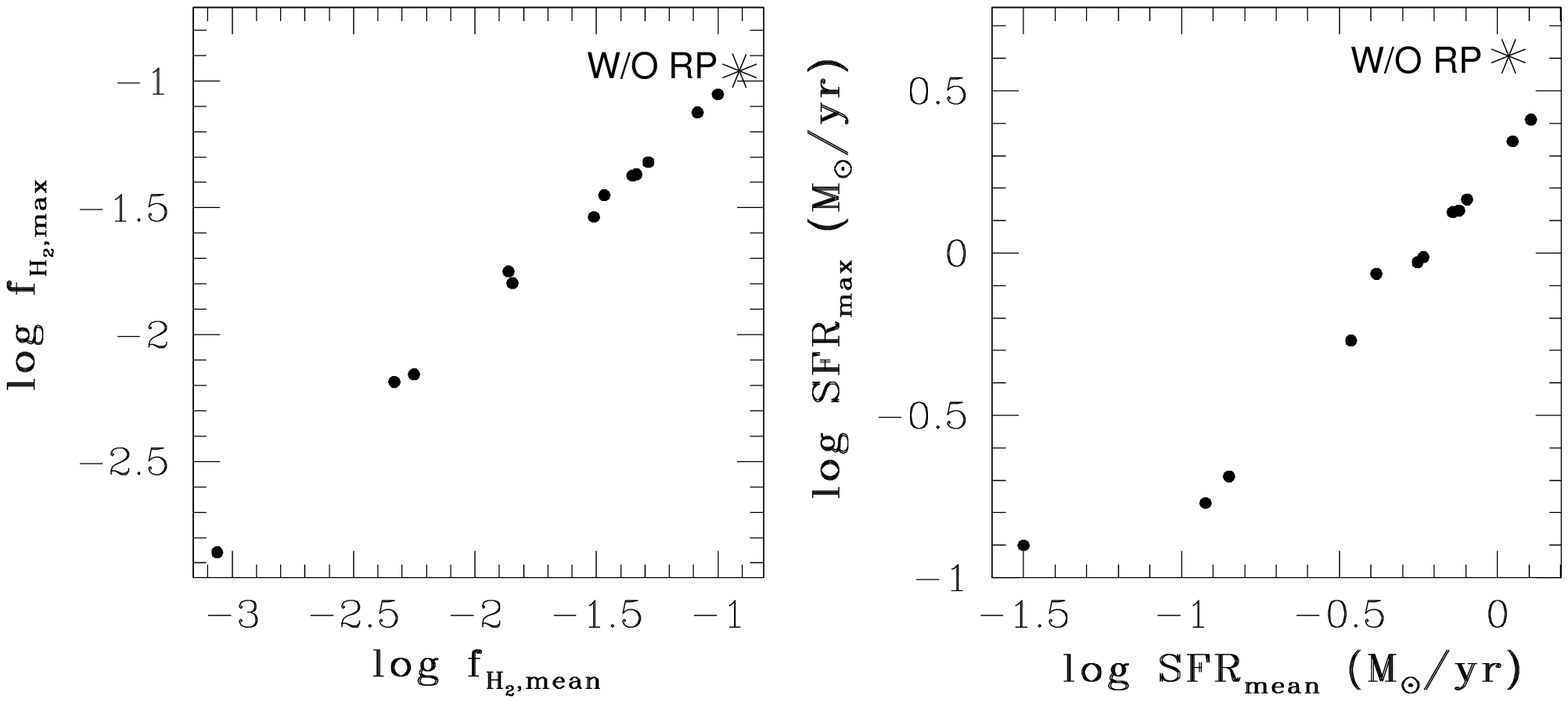,width=14.0cm}
\caption{
The plot of 12 representative models on the $f_{\rm H_2,mean}-f_{\rm H_2,max}$
(left) and ${\rm SFR}_{\rm mean}-{\rm SFR}_{\rm max}$ (right) planes.
Here $f_{\rm H_2,mean}$ and ${\rm SFR}_{\rm mean}$ are 
the mean $f_{\rm H_2}$ and SFR over the 1.1 Gyr evolution, respectively,
whereas
$f_{\rm H_2,max}$ and ${\rm SFR}_{\rm max}$ are the maximum values of
$f_{\rm H_2}$ and SFR, respectively.
For comparison,
the models without dust-related physics is shown as `W/O RP' (without radiation
pressure).
}
\label{Figure. 9}
\end{figure*}

\begin{figure*}
\psfig{file=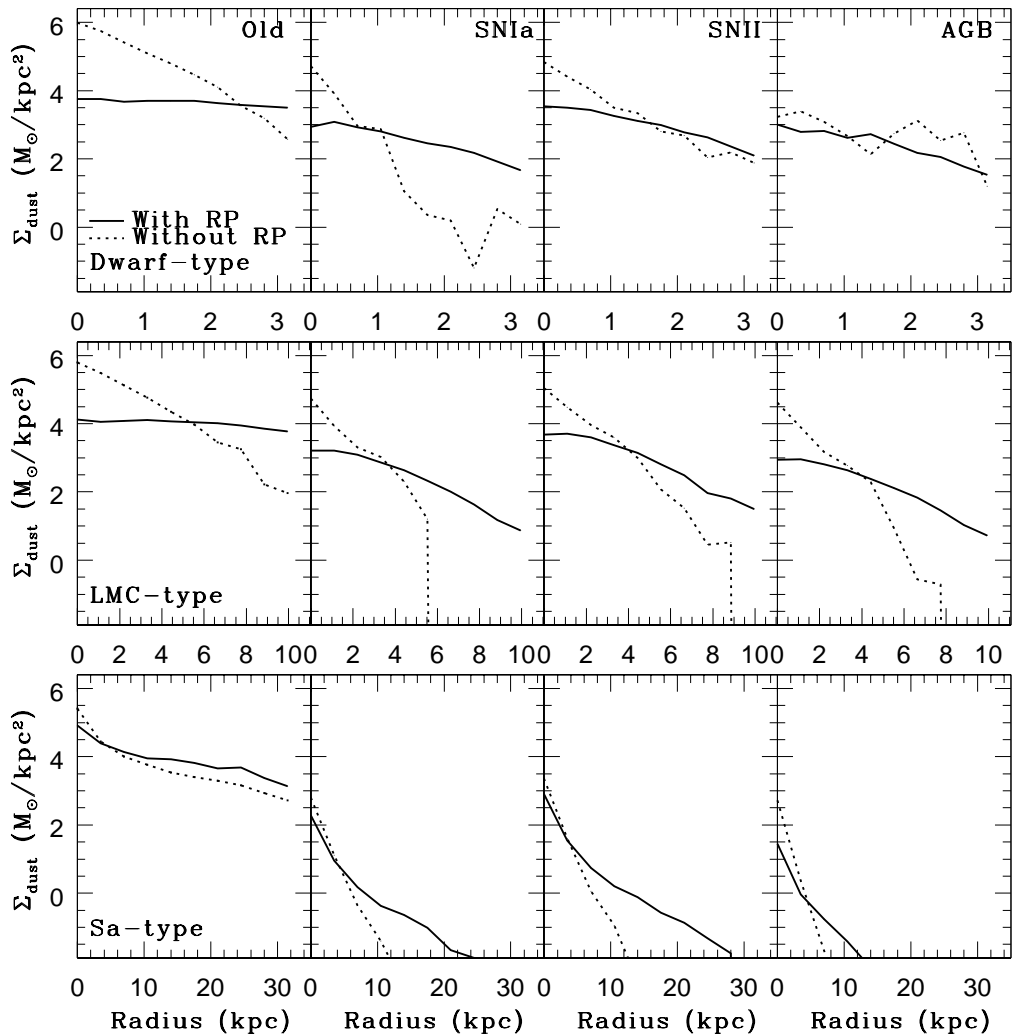,width=14.0cm}
\caption{
The final projected radial density profiles of silicate for old (left),
SNIa (second from the left), SNII (second from the right), and AGB (right)
for the models with (solid) and without (dotted) radiation pressure (RP).
The top, middle, and bottom panels show the results for Dwarf-type,
LMC-type, and Sa-type galaxy models, respectively.
The profile for $R \le R_{\rm g}$, where $R_{\rm g}$ is the initial
gas disk size of a galaxy, is shown for each model.
}
\label{Figure. 10}
\end{figure*}

\begin{figure*}
\psfig{file=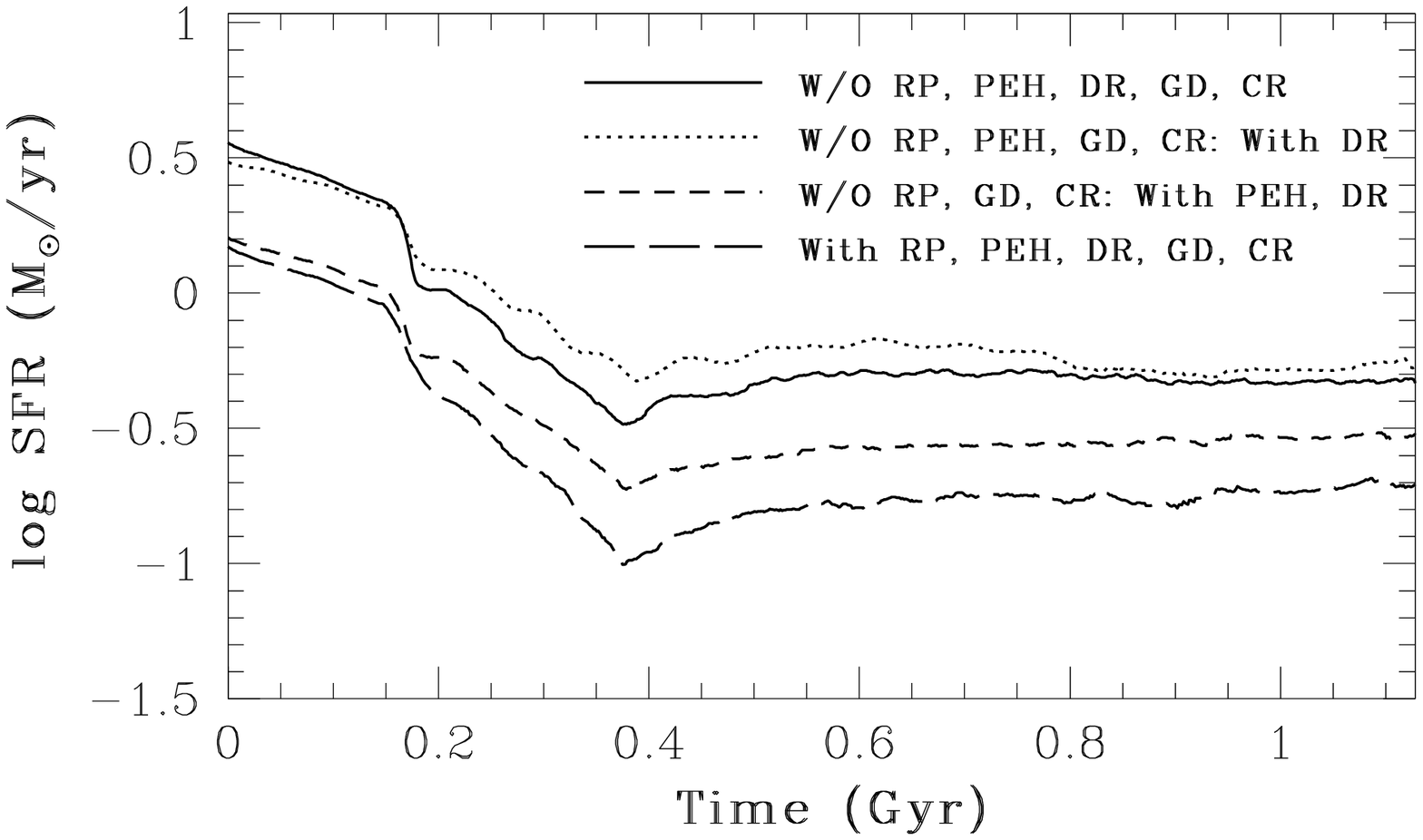,width=14.0cm}
\caption{
The time evolution of SFRs in the MW-type disk models with or without
dust-related physical processes:
without any dust-related processes (solid),
with DR and without ('W/O') RP, PEH, GD, and CR (dotted),
with PEH and DR and without RP, CR, and GD (short-dashed),
and with PEH  DR,  RP, CR, and GD (long-dashed),
}
\label{Figure. 11}
\end{figure*}

\section{Results}

\subsection{MW-type disk}

\subsubsection{Radiation-driven dust evolution}

Fig. 1  shows how the strong radiation field of old stars in the disk
can influence the dynamical evolution of old silicate
in the fiducial MW-type
disk model.  Here `old dust' means dust particles that are initially in
the disk, and therefore the old dust particles do not include
dust produced from SNe and AGB stars after star formation.
Clearly, the inner dust particles can be pushed out by the galaxy-wide
strong radiation
pressure of stars so that a ring-like structure can be once formed
in the disk ($T=0.28$ and 0.56 Gyr). The initially thin disk of old
dust appears to be  thickened owing to the outward transfer
of dust particle by the radiation pressure.
The mean $|z|$ of the dust particles can increase from 0.24 kpc
at $T=0$ Gyr to 1.4 kpc at $T=0.28$ Gyr to 1.9 kpc at $T=1.1$ Gyr
in this model. 
The final distribution of the old dust disk viewed from
edge-on is 
like a disk galaxy with an inner  thick disk ($T=1.1$ Gyr).

Most of the dust particles can not
escape from the galaxy owing to the deep galactic potential of 
this massive disk galaxy ($M_{\rm h}=10^{12} {\rm M}_{\odot}$).
Only 0.4\% of the particles can once locate outside the virial radius
($r_{\rm vir}=245$ kpc) and 98\% of the particles are finally
within $R_{\rm g}$ ($=2R_{\rm s}$)
(=initial gas disk size) in this model.
However, as shown in Fig. 2 for the time evolution of
the  spatial distributions of 
graphite,
a large fraction of graphite appears to escape from the inner 
halo region of the galaxy.
About 19\% of the particles can escape from the galaxy
(i.e., $R > r_{\rm vir}$) within $\sim 1$ Gyr,
and only 24\% of the particles still can be within $R_{\rm g}$
at $T=1.1$ Gyr. 
The mean $|z|$ of the dust particles can increase from 0.24 kpc
at $T=0$ Gyr to 24.5 kpc at $T=0.28$ Gyr to 57.7 kpc at $T=1.1$ Gyr
for the graphite  particles. 
These results confirm the earlier result (F91) that graphite 
can move faster than silicate  and thus
can  escape from galaxies under the strong radiation pressure of stars.

The key factor that can control the dynamical evolution of dust in galaxies
is the ratio ($\beta$) of radiation pressure force to gravitational one:
\begin{equation}
\beta= \frac{ F_{\rm rad} }{ F_{\rm grav} } .
\end{equation}
This $\beta$ parameter is different between silicate and graphite because it depends on
the adopted different dust mass densities and sizes:
\begin{equation}
\beta \propto R_{\rm dust}^2/m_{\rm dust} \propto  R_{\rm dust}^{-1} \rho_{\rm dust}^{-1}.
\end{equation}
Therefore, the $\beta$ value is larger for graphite for the adopted $\rho_{\rm dust}$
(lower for graphite) and $R_{\rm dust}$ (smaller for graphite)
so that graphite can be more strongly influenced by
radiation pressure of stars in galaxies.

As a result of radiation-driven dust transfer, the radial density
profiles of these dust particles can change significantly within
the $\sim 1$ Gyr dynamical evolution of the disk.
Fig. 3 shows that the final distribution of the silicate  particles
at $T=1.1$ Gyr is more flattened than the initial exponential profile adopted
in this model. The final distribution for the inner halo region
$R \sim 20-30$ kpc is more similar to the observed profile of dust
($\Sigma_{\rm dust} \propto R^{-0.8}$) by M\'enard et al 2010 (M10).
The final profile of the graphite  particles 
within the inner halo region is even more
flattened and therefore very similar to the observed flat profile
of dust (M10). These results strongly suggest that the spatial distributions
of many different dust species
can be heavily influenced by radiation pressure of stars in galaxies.
Furthermore, given that our previous simulations without radiation pressure
of stars (B13a) failed to explain the observational results by M10,
these results imply that the origin of the very flat distribution of halo
dust can be closely related to the exertion of stellar radiation pressure on
dust grains.

As shown in Fig. 4,  the newly produced silicate particles
(i.e., SNIa, SNII, and AGB dust)
do not clearly show the flattened density profiles at the selected three
time steps. One reason for this is that  these dust particles are born
in the very thin gas disk (at small $|z|$),  where the dust particles can not be
levitated to a great extent  by stellar radiation pressure
owing to the deeper gravitational potential.
The other reason is that these 'younger' particles have not had
enough time to travel to the outer halo region at $T=1.1$ Gyr.
These results imply that there could be some differences in
the spatial distributions of  dust grains
that were formed in a galaxy at different epochs. 
The mass fraction of silicate that can escape from the galaxy within
$1.1$ Gyr evolution  is 
$0.0007$ for SNIa, $0.0012$ for SNII, and 0.0002 for AGB.
The final mean $|z|$ for SNIa, SNII, and AGB dust are
1.4, 1.5, and 1.3 kpc, respectively.
It is interesting to note that the vertical distribution
of SNII dust is slightly thicker than those of other 
dust types.

The derived outward transport of dust in the disk suggests that radiation
pressure of old disk stars can flatten the radial gradients of $D$.
Furthermore,  this outward transport could be a new mechanism of metal-enrichment
process in the very outer parts of galactic gas disks, if dust can be destroyed
and then returned back to ISM there.
The ejection of dust into the outer halo region of a galaxy means that
both the mean $D$ and dust-to-metal-ratio ($D_{\rm z}$) of the gas disk
of the galaxy can be lowered
significantly. The observed large scatter in $D$ for a given gas-phase
abundance (e.g., Galametz et al. 2011) could be also closely associated
with this dust  ejection process.  Since this radiation-driven dust wind
is not considered in previous one-zone chemical evolution models
with dust (e.g., D98;  Inoue 2003),  $D$ and $D_{\rm z}$ evolution would need to be
re-investigated in future one-zone models for dust evolution.

In the present simulations, the mass of each dust particle ($m_{\rm d}$)
is rather small in comparison with those of other components (DM, stars, and gas).
For example, the initial mass of an old  dust particle in the fiducial MW-type
disk model is only 0.6\% (0.06\%) of the gas (star)  particle mass
($m_{\rm d}=3.6 \times 10^2$ ${\rm M}_{\odot}$). 
The mass-ratio of dust to gas (star)
can also become rather  small for newly formed SNIa, SNII, and AGB dust particle,
because (i) the masses of these dust particles are only small portions of their parent
gas particles and (ii) the significant fraction of dust can be consumed by star formation
and destroyed by SNe.

However,
$m_{\rm d}$ can become much larger than their  initial values
in some local regions owing to dust growth.
Accordingly, the mass range of dust particles can be quite large in a simulation.
For example, $m_{\rm d}$ for SNII dust can range from $3.8 \times 10^{-1}$
${\rm M}_{\odot}$ to $1.9 \times 10^3$ ${\rm M}_{\odot}$ for the fiducial MW disk model.
The average $m_{\rm d}$ for SNII dust produced during $\sim 1$ Gyr evolution
of the MW model is $1.8 \times 10^2$ ${\rm M}_{\odot}$, and the above very low-mass
SNII dust particle is very rare.
In spite of this large mass difference,
the gravitational softening length is fixed at a same value
among all baryonic components.

This means that the dust particles could be 
dynamically heated up (i.e. have more randomized motions)
after they encounter with much more massive stellar and
gaseous (and dark matter) particles. 
Although this possible heating of dust particles can be an undesired numerical artifact
in the present study,  such a dynamical heating (e.g., the formation of very thick
dust disk) can not be clearly  seen in an isolated MW-type disk model. Therefore, it is safe
for us to conclude that the transfer of dust to the outer halo regions shown in Figs. 1
and 2 are due to the effects of radiation pressure on dust  (not due to the numerical
heating caused by particles with vastly different masses).

It should be also noted here that the dust destruction  by warm/hot halo gas in
galaxies is not included in the present simulations. This means that 
the extension of dusty halo caused mainly by radiation pressure
of stars might be overestimated. Dust moving fast in the warm/hot gaseous halo of
a galaxy
might be destroyed efficiently 
(e.g., F91, Bianchi \& Ferrara 2005)
so that the total mass of dust in the galactic halo
can be reduced significantly. As a result of this, the outer profile of the dust
mass density become steeper in the model with dust destruction by the gaseous halo.
This possible effect of gaseous halos of galaxies on dust will need to be investigated
in detail by our future studies.

\subsubsection{The roles of $A_{\rm t}$, $t_0$, and
$Q_{\rm rp}^{\ast}$ in dust evolution and SFH}

Fig. 5 describes how the final projected density profiles of old
dust particles at $T=1.1$ Gyr in the inner halo region
($<30$ kpc) for the MW-type disk model
depend on the adopted $A_t$ (total dust extinction),
dust types (silicate or graphite), initial stellar ages of old disk
stars ($t_0$), and frequency-averaged radiation pressure coefficient
($Q_{\rm pr}^{\ast}$). Firstly, the models with larger $A_{\rm t}$
show steeper radial profiles of silicate,
mainly because 
the higher degrees of dust extinction severely weakens
the stellar radiation pressure on dust so that the spatial distribution
of dust can not change significantly. This dependence can be clearly
seen in the models with graphite, though the distribution
of graphite  is more flattened
than that of silicate for  all models with different $A_{\rm t}$.
Secondly, the models with younger ages of old stars in disks
show flatter density profiles of dust: this is confirmed to be true
for graphite.  The reason for this is that the stronger radiation pressure
of younger stars in a disk can cause the efficient
radial transfer of dust to the outer region of the disk.

Thirdly, the models with larger $Q_{\rm pr}$ show flatter radial profiles
of dust, because dust with larger $Q_{\rm pr}$ can be more strongly
influenced by stellar radiation pressure. The outer radial profiles
of dust and $f_{\rm esc}$
also depend on $A_{\rm t}$, dust types, $t_0$, and $Q_{\rm pr}^{\ast}$. 
For example, $f_{\rm esc}$ for  graphite 
in the MW model with $A_{\rm v}=0.3$ mag
is 0.07, which is about 35\% of the same model with $A_{\rm t}=0$ mag
(shown in Fig 2). 
Fig. 6 describes the time evolution of the spatial distribution
of graphite for the MW model with $A_{\rm t}=0.3$ mag.
In spite of reduced stellar radiation pressure,
some fraction of dust particles still can escape from the galaxy to locate
outside $r_{\rm vir}$ at $T=1.1$ Gyr.
The final slope of the flattened density profile at $R>50$ kpc 
appears to be rather similar to the observed one by M10, which again
suggests that the origin of the observed flat dust distribution
can be caused by radiation-driven dust wind.

Fig. 7 describes the time evolution of the spatial distribution
of silicate for the MW model with a larger dust extinction
($A_{\rm t}=1$ mag) and radiation pressure. Although the outer
distribution of dust is rather flattened owing to the outward transfer of
dust,  the flattening is due largely to the extended
disk of dust. The dusty wind to the vertical direction can not be clearly seen
in this model, because the dust extinction of stellar light by the dusty disk
can weaken the stellar light
and thus prevent the radiation pressure from pushing out dust particles to a large extent.
This derived importance of $A_{\rm t}$
implies that the levels of dust extinction dependent on local properties of ISM   
need to be more self-consistently  modeled so that dust dynamics driven
by stellar radiation pressure in galaxies can be more self-consistently investigated
in our future studies.

Since the dust surface densities of disk galaxies
can change significantly owing to the
radiation-driven dust wind (or dust levitation),
the formation efficiency of ${\rm H_2}$ ($\epsilon_{\rm H_2}$)
on dust grins can also evolve
with time.
The severe reduction of dust mass densities in disks
results in lower $f_{\rm H_2}$
in most models with 
non-zero $Q_{\rm pr}^{\ast}$ in the present study.
The lower $f_{\rm H_2}$ means lower SFRs in the disks, because
the present study adopted the ${\rm H_2}$-dependent SF recipe.
Fig. 8  demonstrate how the SFHs of disk galaxies can be 
influenced by strong
radiation pressure of stars 
in the 12 MW-type disk models with different parameters.
The following five key parameter dependences are found for these
representative models.

Firstly, the models with smaller $A_{\rm t}$ show systematically lower SFRs,
because stronger radiation fields can lower the dust mass densities
of the disks and thus more severely suppress star formation
in ${\rm H_2}$ gas clouds.
Secondly, the models with graphite show lower SFRs than those with silicate,
because graphite can be more strongly influenced by radiation pressure of stars
so that the mass densities of graphite in the disks become lower.
The dependence of SFRs on $A_{\rm t}$ found in the models
with silicate can be clearly seen in the models with graphite.
Thirdly, the models with younger ages of old stars in the disks
show lower SFRs for a given $A_{\rm t}$. The model with 
($t_0,A_{\rm t}$)=(1,0) [Gyr, mag] is not shown in this figure, because
SFR becomes almost 0 in the model even from  the early stage of disk
evolution, because almost all dust is expelled from the disk owing to
the very strong radiation pressure of young stars.
The young disk model with more severe dust extinction, i.e.,
($t_0,A_{\rm t}$)=(1,1) [Gyr, mag], shows lower SFRs in the early stage
of disk evolution ($T<0.6$ Gyr), which implies that star formation
can be severely suppressed by radiation pressure in young disk galaxies at
high $z$.
It is interesting that the model shows higher SFRs at later times
when the radiation
field becomes much weaker and the disk still has a plenty of gas.

Fourthly, SFRs can become lower in the model with $Q_{\rm pr}^{\ast}=2.0$
than $Q_{\rm pr}^{\ast}=0.5$,
because dust is more efficiently removed from the thin disks 
in the model with larger $Q_{\rm pr}^{\ast}$ so that
${\rm H_2}$ mass densities can become lower too.
The model with $Q_{\rm pr}^{\ast}=0$ corresponds to no radiation pressure,
and accordingly
this model shows the highest SFRs at $T<0.2$ Gyr among the 12 models.
Owing to the more rapid ${\rm H_2}$ consumption by star formation at $T<0.2$
Gyr, this model shows lower SFRs in later times.
These results in Fig. 8 clearly demonstrate that (i) radiation pressure on
dust grains can reduce  SFRs in galactic disks  and (ii) the level of this SFR 
reduction (and ${\rm H_2}$ formation efficiencies on dust grains) 
depends on the physical parameters of the disks and dust. 
Fig. 9 summarizes the levels of SFR and $f_{\rm H_2}$ reduction
due to 
stellar radiation pressure on dust grains for the 12 representative models.
Clearly, both the mean and maximum $f_{\rm H_2}$ and SFR can become lower
in the models with radiation pressure.

In real disk galaxies,
their interstellar dust is composed of dust grains with different compositions
and sizes. Furthermore, the ages of old populations in a disk galaxy
should be quite diverse
depending on its formation history
and thus can not be represented by a single age.
Accordingly, the above models with a single $t_0$ and either only silicate
or only graphite would be rather idealized and less realistic.
Therefore it should be stressed here that the 12 models are used
for illustrating the possible effects of radiation pressure on SFHs of 
galaxies. We will discuss how radiation pressure of stars can influence
galaxy-wide SFHs and ${\rm H_2}$ evolution in a more convincing manner
by using a galaxy formation model based on $\Lambda$CDM model in
our forthcoming papers.

\subsection{Other models}

\subsubsection{Low-mass disk  models}

It is confirmed that the influences of stellar radiation pressure
on dust distributions, $f_{\rm H_2}$ evolution, and SFHs derived
for the MW-type disk models are seen also in the LMC-type and Dwarf-type
low-mass disk models. Accordingly, we briefly describe the results
of these models here.  The evolution of dust distributions in these
low-mass disks can be more strongly influenced by radiation pressure
of old stars owing to the shallower gravitational potential. Fig. 10
clearly demonstrates that all dust components (old, SNIa, SNII, and AGB)
in the LMC-type disk model with radiation pressure
show flatter radial density profiles in comparison with
the model without radiation pressure. The old dust have a flatter final
distribution  than other dust components, which is seen in
the MW-type disk model too.

As shown in Fig. 10,
these important roles of radiation-driven dust wind found in the LMC-type disk models
can be clearly  seen in the Dwarf-type disk models,
though the difference of AGB dust distributions
between the models with and without radiation pressure is less clear.
The less clear flattening of dust distribution for AGB dust is due to
the later formation of dust in these models with rather low SFRs.
These results imply that
the present result on the key role of stellar radiation pressure
on the evolution of dust distributions in galaxies can be universal
among galaxies with different masses.
These also suggest that the halos of low-mass disk galaxies can contain
a large amount of dust owing to dust wind driven by stellar radiation
pressure, if the possibly warm halo gas can not destroy the ejected dust
so efficiently.

\subsubsection{Massive bulge models}

Fig. 10 shows that the massive bulge model (Sa-type) with stellar radiation pressure
have flatter density profile of dust for all dust components. However, the slopes
are not so flat as those derived for the LMC- and Dwarf-types disk models,
because the deeper gravitational potential
of the massive bulge can prevent dust from escaping from the thin disk.
Massive bulges in disk galaxies
can suppress the formation of molecular hydrogen in spiral arms (B14),
because they can stabilize the disks against  gravitational instability
(no formation of strong spirals and bars). Therefore, star formation
can be suppressed, and the total amount of dust can not rapidly increase.
The systematically lower dust mass densities in these models are due largely
to the low production rate of dust caused by lower SFRs.
These results combined with those in the MW-type disk models imply
that the 3D distributions of dust in galactic halos could be different
in galaxies with different Hubble types.

\subsubsection{Other dust effects}

We have so far focused on the results of the models that do not
include dust-related physics (e.g., photo-electric heating
and dust destruction by SNe) other than
stellar radiation pressure on dust grains. This is mainly because
our main purpose is to investigate how the stellar radiation pressure,
which is not included in B13a, can possibly influence galaxy evolution.
The influences of other each dust-related physics on galaxy evolution
will be separately discussed in our  forthcoming papers.
Although photo-electric heating (PEH), cosmic-ray heating (CR), dust-gas heating
(DG),  gaseous drag of dust (DR), and dust destruction
can influence the thermal history of gas and dust, they can not 
change the evolution of the 3D dust distribution so dramatically than
stellar radiation pressure (RP): this is indeed confirmed in the present
study.

The essential role of stellar radiation pressure on dust evolution
can be seen in the models with RP, PEH, CR, DR, DG, and dust destruction.
The suppression of ${\rm H_2}$ formation and star formation by radiation
pressure of stars can be seen
in these models, though other dust-related physics can significantly
modify the suppression effects.
Fig. 11 shows a number examples
of key effects of these dust-related physics
on SFHs of disk galaxies.  The gaseous drag (DR) 
in the model without RP yet with DR can prevent dust particles
from escaping from the thin gas disk where star formation can occur so that
the SFR can be slightly higher in the model than in the model without RP and DR.
The model with PEH shows a systematically low SFR owing to the increased
gaseous temperature (in particular for $T<10^4$ K) of cold ISM.
This suppression of SF by PEH can be seen in the models with or without
RP, which implies that PEH can play a key role in SFHs of galaxies.
This role of PEH should be extensively explored in our future works.

As shown in Fig. 11, the model with all dust-related physical processes
shows the lowest SFR among the four models.
This implies that if numerical simulations of galaxy formation
and evolution do not include  dust physics properly,
then they could severely overestimate the
galaxy-wide SFRs.
 Given that such dust effects on galactic SFHs
can be stronger in galaxies with lower masses,
the results in Fig. 11 implies that the total stellar masses
of low-mass disk galaxies 
could have been overestimated 
in the previous simulations of galaxy formation without dust.
It should be stressed that these discussion on galactic SFHs is based
on the adopted ${\rm H_2}$-dependent SF models. If SFRs are simply
proportional to total gas densities
(neutral + molecular hydrogen), then the influences of
dust-related physics on SFHs could be reduced to a large extent,
because SFRs do not depend on the ${\rm H_2}$ evolution that is strongly
influenced by dust evolution.

\begin{figure*}
\psfig{file=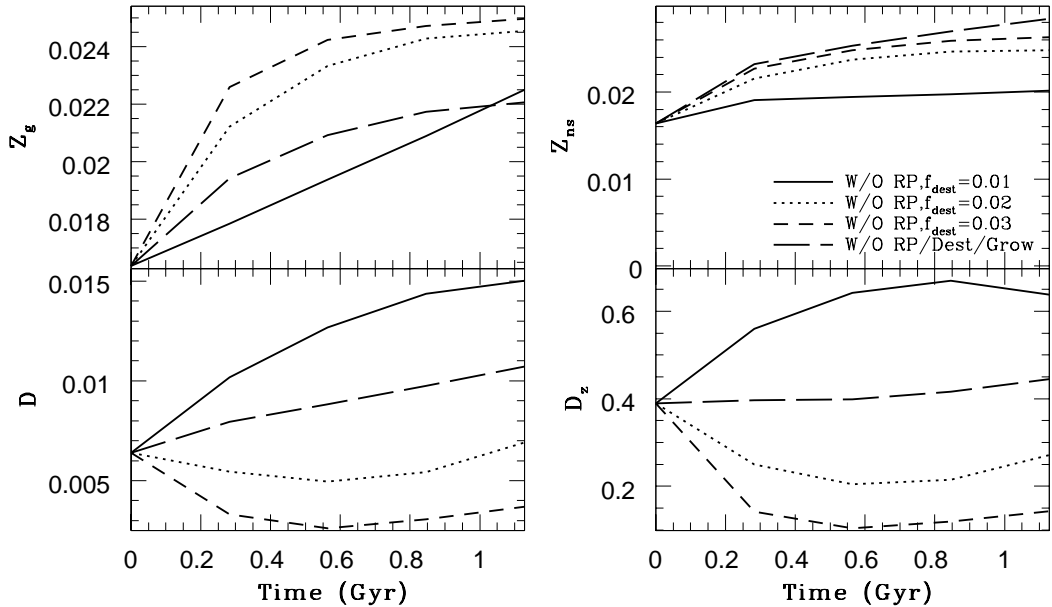,width=14.0cm}
\caption{
The time evolution of mean gas-phase metallicity ($Z_{\rm g}$, upper left),
mean metallicity of new stars ($Z_{\rm ns}$, upper right),
dust-to-gas-ratio ($D$, lower left), and
dust-to-metal-ratio ($D_{\rm Z}$, lower right) for the four MW-type disk
without  stellar radiation pressure. Silicate is adopted for the 
dust component.
The solid, dotted, and short-dashed lines indicate 
the models with different $f_{\rm dust}=0.01$, 0.02, and 0.03, 
respectively. For comparison, the results for the model without
dust growth and destruction are shown by a long-dashed line.
Since there is no new star at $T=0$, the initial $Z_{\rm g}$ value
is plotted for $Z_{\rm ns}$ at $T=0$.
}
\label{Figure. 12}
\end{figure*}

\begin{figure*}
\psfig{file=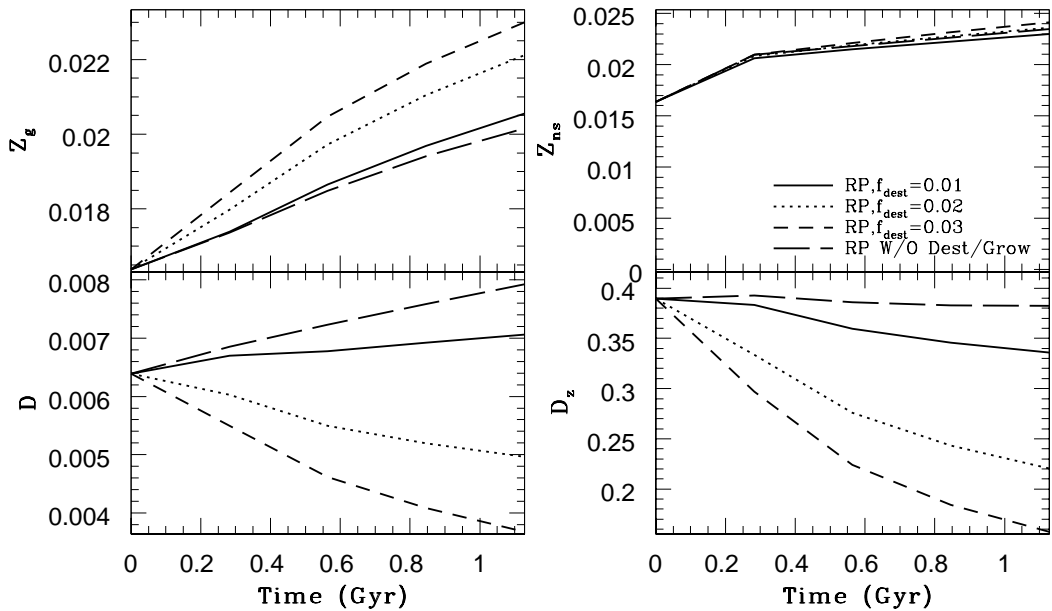,width=14.0cm}
\caption{
The same as Fig. 12 but for the MW-type disk model with stellar radiation
pressure.
}
\label{Figure. 13}
\end{figure*}

\section{Discussion}

\subsection{How can galactic chemical evolution be possibly influenced by dust 
evolution?}

The chemical evolution of galaxies can be influenced largely by
their SFHs that are controlled by  a number of physical processes,
such as merging histories of galaxies, stellar winds by
energetic feedback effects, and  molecular cloud formation.
Given that dust grains are the major formation sites of ${\rm H_2}$ gas
in galaxies,
the evolution of dust abundances, compositions, and sizes can be 
a key factor for ${\rm H_2}$ evolution and thus for chemical evolution
of  galaxies. Although a full investigation of this issue should be done in our
future studies with the present new code,
it would be instructive for the present study to discuss the possible
influences of dust evolution in galactic chemical evolution briefly.
Fig. 12 shows the time evolution of gas-phase metal abundances ($Z_{\rm g}$),
metal abundances of new  stars ($Z_{\rm ns}$), dust-to-gas-ratios ($D$),
and dust-to-metal-ratios ($D_{\rm z}$) for the MW-type models without
radiation pressure (`W/O RP') and different $f_{\rm dest}$. 
For comparison, the results of the model without dust growth and destruction
are shown in this figure.

Clearly, the evolution of $Z_{\rm g}$, $Z_{\rm ns}$,
$D$, and $D_{\rm z}$  depends strongly
on how much fraction of dust can be destroyed by SNe in a star-forming cloud.
In the model with $f_{\rm dest}=0.03$,  a larger fraction of dust can be
returned back to ISM as gas-phase metals owing to the more efficient
dust destruction  so that gas-phase metallicity
($Z_{\rm g}$) can increase more rapidly and to a larger extent.
Because of the higher gas-phase metallicity of the ISM,
the new stars can also have higher metallicities ($Z_{\rm ns}$).
As a result of this, $D$ and $D_{\rm z}$ can decrease with time in this model
with higher $f_{\rm dest}$. On the other hand, the model with 
low $f_{\rm dest}$  (=0.01) shows slower increase in $Z_{\rm g}$ and 
$Z_{\rm ns}$ and significant increase in $D$ and $D_{\rm z}$.
These results suggest that dust destruction by SNe in ISM can greatly influence
the chemical evolution of galaxies.

As shown in the present study,
the stellar radiation pressure can levitate dust to the halo regions of
galaxies where gas densities are so low that dust can not efficiently
grow. The radiation pressure also can suppress the galaxy-wide star formation
and consequently slow down the chemical enrichment processes of galaxies.
Therefore,  the stellar radiation pressure on dust can influence
the time evolution of $Z_{\rm g}$, $Z_{\rm ns}$, $D$, and $D_{\rm z}$.
Fig. 13 clearly demonstrates that $Z_{\rm g}$ and $Z_{\rm ns}$ can more slowly
increase with time in the models with RP.
Although the model with  a low $f_{\rm dust}$ (=0.01) shows a slow increase
in $D$, it shows a slight  decrease of $D_{\rm z}$.
In this model, gas-phase metallicity can steadily increase due to star formation,
because ISM can not be so efficiently ejected into the halo region.
However, dust-phase metallicity
can not increase so efficiently through accretion of
gas-phase metals onto dust grains, because a significant fraction
of dust can be ejected into the halo region where gas density is low
Therefore, the destruction of dust by SNe can lower $D_{\rm z}$ 
even in this model with low $f_{\rm dest}$.

Thus,  these results suggest that we need to carefully model 
the dust-related physics in ISM of galaxies in order to discuss
the chemical evolution of galaxies. One of the most important parameters
in the present new model is $f_{\rm dest}$, which needs to be 
determined by another sets of pc-scale simulations of forming and
evolving molecular clouds with young stars and SNe. Since
$f_{\rm dest}$ is defined for an entire star-forming cloud with
multiple SNe,  we can not simply use the results of simulations 
for dust destruction by a single SN event (e.g., Nozawa et al. 2003).
Given that $f_{\rm dest}$ can so strongly influence the time evolution
of dust and metal abundances, it is doubtlessly our future study
to determine the reasonable range of $f_{\rm dest}$ for galaxy-scale
simulations. The best way to do so  would be to compare
the observed scaling relation of dust properties
(such as a $D$-$Z_{\rm g}$ relation, e.g., Galametz et al. 2011)
the corresponding simulations with different $f_{\rm dest}$.
It would be also necessary for our future theoretical works to
search for the possible range of $f_{\rm dest}$ and its dependence
on ISM properties based on sub-pc-scale simulations of star-forming 
molecular clouds.

\subsection{Galaxy life slowed down by dust}

The present study has first shown that the radiation pressure of stars
on dust can suppress the galaxy-wide star formation,
because the pressure can  reduce the ${\rm H_2}$ mass fractions that can
control SFRs of galaxies. 
This suggests that radiation-driven dust wind can change the chemical
evolution of disk galaxies not only through the removal of some metals 
in the disks (Bekki \& Tsujimoto 2014)
but also through `slowing down' the gas consumption by star formation
thus the chemical enrichment processes. 
Accordingly, the present study implies that
this `slowed down' lives of galaxies could be one of key results of
galaxy formation and evolution influenced by interstellar dust.
The present study also suggests that if numerical simulations of galaxy formation
and evolution do not include  dust-related physical processes in a self-consistent manner,
then they
are likely to over-predict the total gas mass converted into new
stars.

The present study, however, has shown that the level of SF suppression
by radiation-driven dust wind depends on the adopted constant dust extinction
of ISM and the dust composition.  This implies that
our future more sophisticated models with a self-consistent model
for radiative-transfer of stellar light in dusty ISM (which
can properly predict $A_{\rm t}$ variation in time and space)
would produce different results.
Furthermore, the SF suppression
can be clearly seen in the present simulations, because
SFRs are assumed to depend on  the mass densities of ${\rm H_2}$
the formation efficiency of which depends strongly on dust densities.
Accordingly, if SFRs depend on the total gas densities rather than
${\rm H_2}$ densities, then the SF suppression would not be so clear
as the present models have shown.
Thus, the level of SF suppression by radiation-driven dust wind needs
to be investigated by our more sophisticated models with a fully
self-consistent treatment of stellar light influenced by local dust.

The possible influences of dust on galaxy formation and evolution
revealed by the present study are summarized in Table 6.
Given that the present new model is still idealized and less realistic
in some points (e.g., non-inclusion of dust size distributions),
these possible influences might be modified (at least in a quantitative
sense) in more sophisticated models for dust formation and evolution
in galaxies.
Furthermore, the present simulations are only for disk galaxies in an  isolated
environment without no dynamical and hydrodynamical interaction with
other galaxies.  Some unknown influences of dust on galaxy formation
would be likely to be found in numerical simulations of galaxy formation.
Therefore, such influences of dust  need to be re-investigated thoroughly
in future numerical simulations of galaxy formation with a self-consistent
modeling of dust-related physical processes in ISM.

\begin{table*}
\centering
\begin{minipage}{160mm}
\caption{A list of possible influences of dust on galaxy formation
and evolution based on the present results. These need to be re-investigated
thoroughly in future theoretical studies based on numerical simulations
of galaxy formation and evolution with dust-related physics.} 
\begin{tabular}{lll}
Dust-related physical processes  &  Galactic properties  & Possible influences \\
 &  & \\
Radiation pressure on dust (RP) &  Dust distribution &  
Flattening of radial profiles and formation of outer dusty halos \\
 & Gas dynamics &  Less efficient ${\rm H_2}$ cooling in ISM with lower $T_{\rm g}$ \\
 & Star formation &  Suppression of galaxy-wide star formation \\
 & ${\rm H_2}$ content &  Reduction in thin gas disks \\
 & Chemical enrichment &  Slowed down metal-enrichment\\
Photo-electric heating (PEH) &  
Star formation & Significant suppression of galaxy-wide
star formation  \\
 & ${\rm H_2}$ content &  Reduction in thin gas disks \\
Gas-dust drag (DR)  & Dust distribution &  
Prevention of dust from escaping from galaxies \\
 &  Star formation & Mitigation of the RP's suppression effects of star formation \\
\end{tabular}
\end{minipage}
\end{table*}

\subsection{Future directions of the new live dust particle method}

The present new four-component chemodynamical code will enable us to investigate
various aspects of dust-affected
galaxy formation and evolution across the comic time. The proposed
live dust particle method will make it possible for us to discuss the different
distributions of gas and dust in galaxies,  the cosmic evolution of dust sizes
and compositions, and the abundances of interstellar matters formed on dust grains.
Below, we briefly discuss what we can investigate by using the new simulation code
(The figure in Appendix C
would be useful for readers of this paper to understand
the following discussion better).

\subsubsection{Dust-gas decoupling}

B15 failed to reproduce the observed flat profiles of dust halos of galaxies  (M10),
because the model adopted in B15 did not include dust-gas decoupling processes in ISM
(e.g., radiation pressure and dust-gas drag).
Furthermore, the observed rather high dust-to-gas-ratio ($D \sim 0.05$,
which is about 6 times larger than the solar neighborhood) in the outer halo of M81
by Xilouris et al. (2006) was very difficult to be explained by B15 for the same reason.
As shown in this paper, the radial density profile of dust in a galaxy can be flattened
by the effects of radiation pressure of stars (thus the outer halo can have
higher $D$).  Therefore, it is doubtlessly worthwhile
for our future studies to investigate whether such an important effect of radiation
pressure can be clearly achieved in cosmological hydrodynamical simulations of 
galaxy formation.

\subsubsection{Dust size distribution}

The present chemodynamical model includes the six key dust-related physical
processes of galaxies, i.e., dust formation, mixing,  growth,  destruction,
consumption, and catalysis.  The formation of
${\rm H_2}$ on dust grains is one of important interstellar chemistry
processes on dust grains ('catalysis') that can be discussed in the present study.
The present new live dust particle method combined with the proposed
four-component galaxy model (DM+stars+gas+dust)
can predict dust and gas 3D distributions separately and  distributions
and kinematics of different dust species (e.g., AGB and SN dust) separately
too.
However, the live particle method is still less realistic in modeling
some aspects of dusty ISM, such as the size evolution of dust 
and the coagulation process
of dust.

The bottleneck of the new method is that the total number of
dust particles can endlessly increase if the formation of dust with different
sizes and components in star-forming regions and evolved stars 
is considered in a fully self-consistent manner. This means that
the CPU/GPU time required for the calculation of dust physics at each time step
of a simulation can become 
progressively longer as the simulation proceeds. 
The total number of dust particles ejected from a single stellar particle
($N_{\rm ej}$) is 3 in the present study so that the total number of particles
($N$) can not dramatically increase. However, if we investigate the time
evolution of dust sizes,  $N_{\rm ej}$  would possibly need to be
more than $\sim 100$ to represent a statistically enough sample
of dust grains with different sizes.
This means that we need to develop a fast algorism for calculating
the dust-related physics of ISM  in a simulation in order to consider the 
dust size evolution in a  self-consistent manner.

\subsubsection{Self-consistent SED construction}

Constructing the SEDs for arbitrary 3D geometries of gas, dust, and stars
in galaxies based on theoretical and numerical
models of galaxy formation has been one of key ares of
theoretical studies of galaxies
(e.g., Bekki \& Shioya 2000, 2001;  Jonsson 2006; Popescu \& Tuffs 2013).
These simulated SEDs and those based on some phenomenological models (e.g., Dale et al. 2001)
are very useful in interpreting the observed energy budgets (e.g.,
fraction of infrared light)  of dusty star-forming galaxies. 
In order to estimate SEDs from SSPs (e.g., Bruzual \& Charlot 2003), 
the previous studies needed to assume (i) dust-to-metal-ratios
(to drive dust abundances from gas metallicities)  and (ii) 
dust size distribution (for deriving the dust extinction curves).
If our future simulations
with the new live dust particle method can predict dust size distributions properly,
then they do not have to adopt the above assumption (ii) (the above (i) is explicitly
derived in the present code already). Thus, the self-consistent construction
of SEDs based on dust size evolution will be possible in our future studies.

\subsubsection{Dust-corrected cooling}

Although the gas-phase metallicity (that does not include dust-phase metal)
of a gas particle
should be used for estimating the radiation cooling
rate of the gas in numerical simulations of galaxy formation and evolution,
the total ISM metallicity (i.e., gas-phase + dust-phase metal) has been used
in almost all simulations for cooling estimation,  because dust was not included
in the simulations.  Although the present new model properly considered this
dust-corrected cooling process, it did not include the possible time-dependent
IMF, which can cause significant changes in dust evolution,
star formation histories, and chemical evolution of galaxies
(e.g., Bekki 2013b; Recchi \& Kroupa 2015). 

Furthermore, the present study
did not consider the radiative cooling rates that depend on both [Fe/H]
on [Mg/Fe] (De Rijcke et al. 2013) thus on the detailed chemical evolution of galaxies.
Given that the dust-depletion levels are quite different between different elements
(e.g., Mg, Fe, and Ca), the present results might be changed to some extent if
the cooling rates by De Rijcke et al. (2013) are adopted.
Thus our future studies will need to include both temporal and spatial variations
of IMFs and more sophisticated cooling models (De Rijcke et al. 2013) in order
to model the thermal and dynamical evolution of ISM regulated dust in a more
self-consistent way.

\subsubsection{Dust destruction by multiple SNe}

The present study has introduced a new parameter for the mass fraction of dust
destroyed by SNe ($f_{\rm dest}$). In the simulation code,
the dust particles that are within less than $R_{\rm dest}$ ('destruction
radius') from a SN are assumed to be destroyed and consequently lose some
fraction ($f_{\rm dest}$) of dust. Since a cluster of stars is formed from
a gas particle, the previous results based on a model for dust destruction by a single SN
can not be simply used in the present model. 
Given that $f_{\rm dest}$ can control chemical evolution of galaxies,
the possible range of $f_{\rm dest}$ in star-forming gas clouds
with multiple SNe exploding at different epochs 
needs to be investigated in our future simulations.

The mixing radius ($R_{\rm mix}$) of ejected metals from SNe and AGB stars
and the dust destruction radius ($R_{\rm dest}$) are set to be the same in the present
study. The appropriate values for $R_{\rm mix}$ and $R_{\rm dest}$ are determined 
from the adopted spatial resolution of a simulation (i.e., from $\epsilon_{\rm g}$).
However, these two could be significantly different, because dust destruction
processes might be more sensitive to physical properties of ISM (e.g., magnetic
fields and gas shock speed;  Jones et al. 1994).  This means that we need to perform
high-resolution sub-pc scale numerical simulations on the dynamical evolution
of dust and metals ejected from SNe and AGB stars in galactic potentials.
The outputs from such simulations will be able to be included in larger galaxy-scale
simulations so that we can investigate the dust-regulated galaxy formation and evolution
for reasonable ranges of $R_{\rm mix}$ and $R_{\rm dest}$.

\subsubsection{Physical properties 
of interstellar ice and organic matter}

Although we have so far focused on the formation of ${\rm H_2}$ on dust grains,
the surface of grains in ISM is crucial for the formation and evolution
of many different interstellar species (as `catalysis'), such as ice and organic matter
(e.g., Herbst \& van Dishoeck 2009). Accordingly, the present new
dust particle methods will enable us to predict the formation efficiency
of many different species other than ${\rm H_2}$ by considering physical
conditions of gas and dust and chemistry on dust grains. 
As a dust particle moves in ISM of a galaxy, the dust temperature,
the collision rates of atoms and molecules,  the strength of ISRF,
and other physical properties of dust and ISM can change. If these time
evolution of the physical properties of the dust and its environments
can be investigated by our future simulations, we will be able to 
predict the time evolution of interstellar species (e.g., organic matter).
Given that recent infrared observations have already shown different abundances
of ${\rm H_2O}$ and ${\rm CO_2}$ ices in different galaxies 
(e.g., Shimonishi et al. 2010),  this prediction is  something that
we should do not in far future but in near future, though the modeling
of ${\rm H_2O}$ and ${\rm CO_2}$ ices might be a formidable task.

\section{Conclusions}

We have developed a new four-component model of galaxy evolution
in which a galaxy consists of dark matter, stars, gas, and dust.
The dust component is represented by 'live' dust particles so that
the present new model enables us to investigate separately the evolution
of gas and dust.The dust particles are 'live' in the following three senses.
Firstly,  dust particles can gravitationally interact
not only  with other components (e.g., dark matter) but also with
other dust particles. Secondly,  dust can be destroyed by SNe and reduce
significantly its mass, and also it can grow by accretion of gas-phase
metals in ISM.  Thirdly,  dust grains can be the formation sites
of water, organic compounds, and amino acid closely related to the origin
of life, and accordingly
we can study the evolution of these interstellar matter  essential for life
by using our new simulations in future.

First we  have described the details of the model and discussed
the limitations and advantages of the model in investigating galaxy
formation and evolution. We then have  investigated 
the possible roles of dust in galaxy evolution by using
the new code and obtained some
preliminary results. In particular, we have tried to
grasp the possible roles of stellar radiation pressure on dust grains 
in the evolution of dust,  gas-phase metals, and star formation
histories (SFHs) of galaxies.
The preliminary results that we have obtained
in this study are briefly summarized as follows:  \\

(1) The radiation pressure of stars on dust in disk galaxies can influence
the time evolution of 3D dust distributions. Dust particles in a disk galaxy
can be levitated from the initially thin disk and some fractions of dust
can escape from the disk to locate outside the virial radius of its
dark matter halo. This dynamical evolution of dust derived by stellar
radiation strongly depends on the model parameters such as stellar ages
($t_0$), total dust extinction ($A_{\rm t}$), dust composition
(e.g., silicate or graphite), and frequency-averaged
radiation pressure coefficient ($Q_{\rm pr}^{\ast}$). \\

(2) The two basic roles of stellar radiation in the time evolution
of 3D dust distributions are (i) thickening of the vertical structure
of dust and (ii) flattening of the radial density profile.
The above four parameters (e.g., $A_{\rm t}$)
can determine the degrees of these thickening
and flattening in the mass distributions of dust.
For example, the models with higher $A_{\rm t}$ show a smaller
degree of flattening
in the dust distribution whereas those with younger disk ages
(i.e., smaller $t_0$) show a larger degree of the flattening. \\

(3) Stellar radiation pressure can reduce the mass density of dust
in a galactic disk so that it can also reduce the ${\rm H_2}$ formation
efficiency on dust grains. As a result of this,
the star formation rate (SFR) can be severely reduced with the
reduction level depending largely on $A_{\rm t}$.
It should be noted, however, that this suppression of star formation
by stellar radiation pressure is due to the adopted assumption
that SFRs depend on the mass density of ${\rm H_2}$ gas 
(not the total gas density) in the present
study. Therefore, it is possible that the level of SF suppression
by radiation pressure is overestimated in the present study. \\

(4) The inclusion of other dust-related physical processes,
such as photo-electric heating and gas-dust heating, does not
change the present results on the roles of stellar radiation pressure
in galaxy evolution. It is found that photo-electric heating can 
suppress galaxy-wide star formation in the models with and without
radiation pressure. Furthermore, gaseous drag of dust can lead to
the slight enhancement of galactic SFRs, because the drag can prevent
dust from escaping from thin galactic disks.
We need to confirm these interesting effects of
dust on galactic SFHs in our future studies by exploring a much wider
parameter space for dust-related physical processes. \\

(5) One of the most important parameters for the evolution of
dust and metal abundances is $f_{\rm dest}$, which  is defined as
the mass fraction of interstellar dust that is destroyed by SNe
in a star-forming gas cloud. Although a possibly reasonable value 
of $f_{\rm dest}$ is $\sim 0.01$ in the present study,
this parameter is yet to be constrained so well by observational
and theoretical studies.  Given its importance, it is our future
study to determine the possible range of $f_{\rm dest}$ and its
dependence on the physical properties of star-forming gas clouds. \\

(6) The present results suggest that although dust does not influence
galaxy evolution so dramatically as SN and AGN feedback effects do,
it can play significant roles in many aspects of galaxy formation and evolution.
For example,
the removal of  metals through dust wind,  suppression of ${\rm H_2}$ formation
due to  dust removal,  and photo-electric heating of gas by dust in ISM
can influence chemical evolution,
star formation histories,
and stellar and gas dynamics in galaxies, respectively.
These roles of dust in galaxy formation and evolution will need to be investigated
extensively in future theoretical studies of galaxies for better understanding the formation
and evolution processes of galaxies. \\

Thus the present new live dust particle method has enabled us to
reveal the possible influences of dust-related physical processes
on galaxy evolution. 
The dust properties of galaxies have been derived in the
'post-processing' of simulation data for almost all previous theoretical simulations
of galaxy formation and evolution (e.g., derivation of dust mass from metal mass
by assuming dust-to-metal ratios). The present study suggests
that this 'post-processing' approach might not be realistic
because dust itself can cause significant changes in galaxy evolution processes.

The present new model is still somewhat idealized 
(e.g., non-inclusion of dust size distributions in ISM) and therefore
needs to be further improved so that we can properly investigate
the physical properties of dust (e.g., evolving dust compositions
and sizes) as well as their effects on galaxy formation and evolution,
in particular, the evolution of galactic SEDs and molecular gas fraction.
In our next paper, we will discuss how galaxy formation processes can be 
influenced by dust in a $\Lambda$CDM cosmology by using a more sophisticated  
live dust particle model.

\section{Acknowledgment}
I (Kenji Bekki; KB) am   grateful to the referee  for  constructive and
useful comments that improved this paper.
Numerical simulations  reported here were carried out on
the three GPU clusters,  Pleiades, Fornax,
and gSTAR kindly made available by International Center for radio astronomy research(ICRAR) at  The University of Western Australia,
iVEC,  and the Center for Astrophysics and Supercomputing
in the Swinburne University, respectively.
This research was supported by resources awarded under the Astronomy Australia Ltd'
s ASTAC scheme on Swinburne with support from the Australian government. gSTAR is funded by Swinburne and the Australian Government's
Education Investment Fund.
KB acknowledges the financial support of the Australian Research Council
throughout the course of this work.

\appendix

\section{Required number of dust particles}

\begin{figure}
\psfig{file=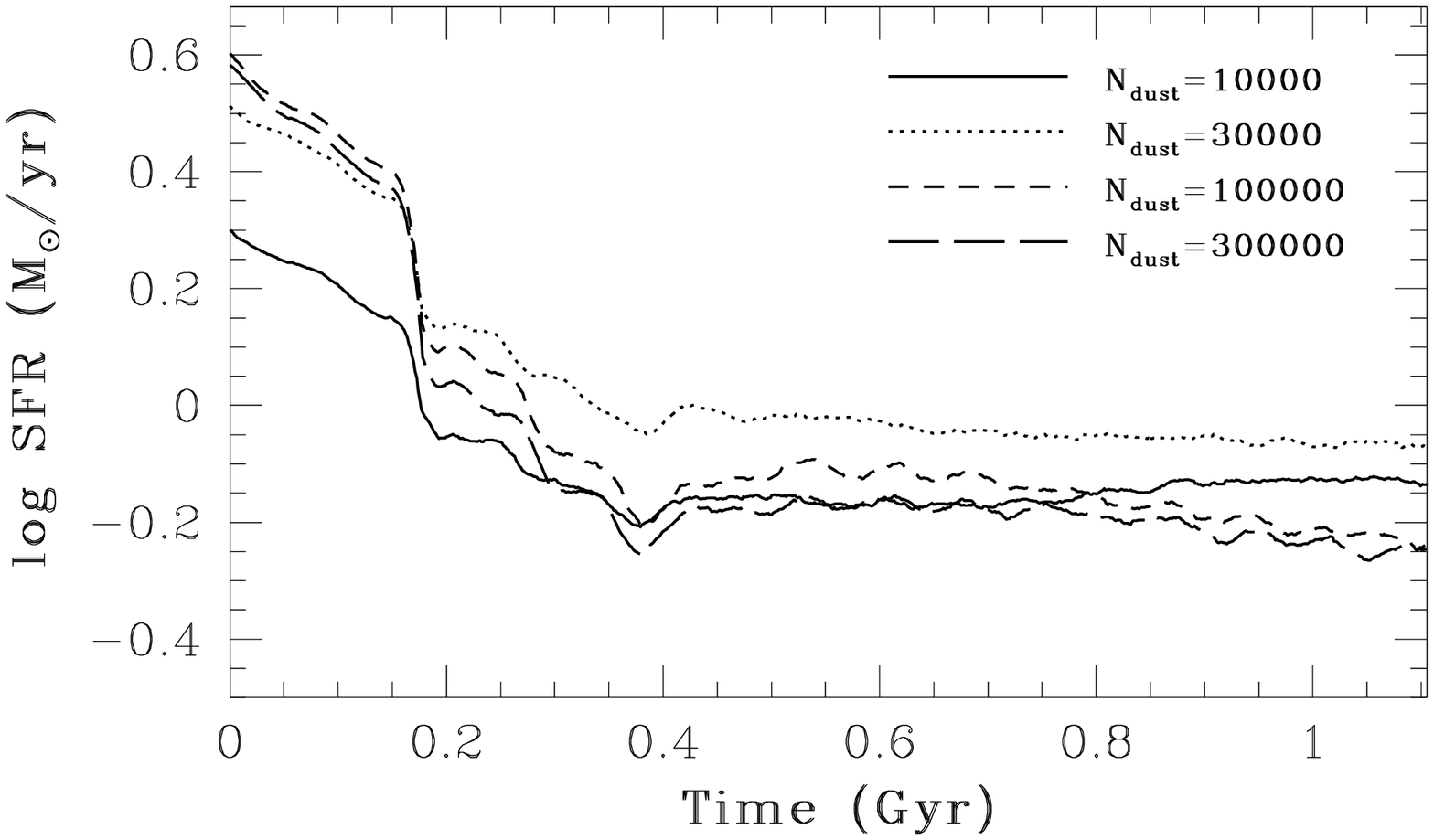,width=8.0cm}
\caption{
The star formation histories of the fiducial MW-type disk models
with $N_{\rm dust}=10000$ (solid), 30000 (dotted), 100000 (short-dashed),
and 300000 (long-dashed).
}
\label{Figure. A1}
\end{figure}

In the present live dust particle method,
the formation of ${\rm H_2}$ is possible only on the surface of dust grains
in galactic ISM.  Accordingly, the time evolution of $f_{\rm H_2}$ and 
SFHs of galaxies depend on the dynamical evolution of dust. It is therefore
possible that the present results on $f_{\rm H_2}$ evolution and  SFHs of 
galaxies depend on the total number of dust particles ($N_{\rm dust}$)
adopted in the present simulations. In order to investigate this important
issue,  we have run four comparative MW-type disk models without
stellar radiation pressure and 
different $N_{\rm dust}$ and investigated whether the results depend on
$N_{\rm dust}$.

Fig. A1 shows that the SFRs in MW-type disk galaxies do not depend 
strongly on $N_{\rm dust}$ as long as $N_{\rm dust} > 30000$ 
(30\% of the total gas particle number). The model with $N_{\rm dust}=10000$
appears to underestimate the SFR in the initial bursty star formation 
in the central region of the disk. It is confirmed that
the final spatial distributions of dust does not depend strongly
on $N_{\rm dust}$ either. This weak dependence on $N_{\rm dust}$
is very encouraging, because we do not have to adopt an excessively
large number of dust particles to simulate the dust evolution of galaxies.
Thus we conclude that the adopted number of $N_{\rm dust}$ is enough
to investigate the possible roles of dust in galaxy evolution.

\section{Dependence on $R_{\rm grow}$}

\begin{figure}
\psfig{file=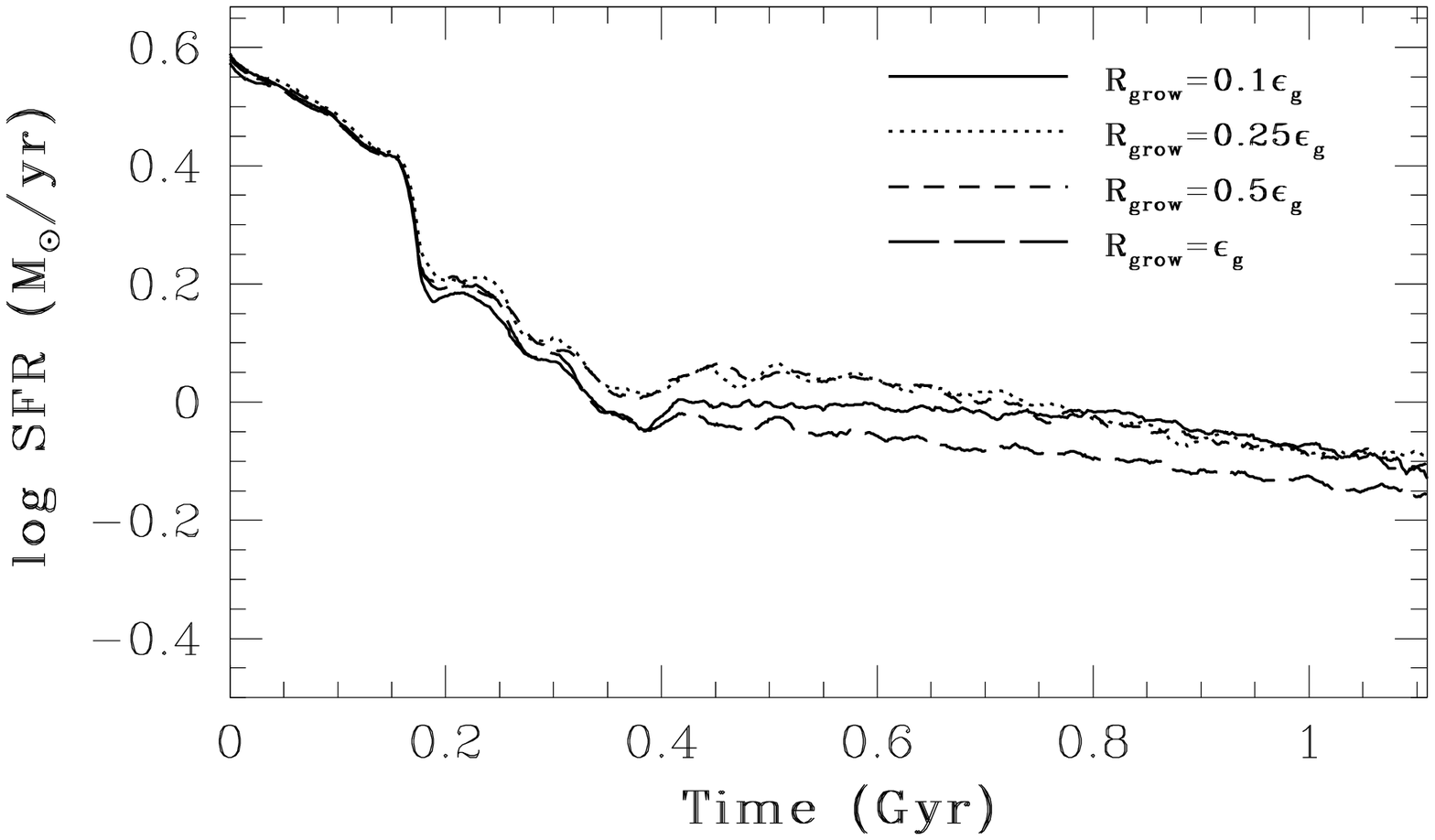,width=8.0cm}
\caption{
The star formation histories of the fiducial MW-type disk models
$R_{\rm grow}=0.1 \epsilon_{\rm g}$ (solid),
$0.25 \epsilon_{\rm g}$ (dotted),
$0.5 \epsilon_{\rm g}$ (short-dashed),
and $\epsilon_{\rm g}$ (long-dashed),
}
\label{Figure. B1}
\end{figure}

Dust growth via accretion of gas-phase metals onto already existing dust
grains and and dust destruction by SNe are both assumed to proceed locally
depending on the physical conditions of ISM 
in the present study.  In order to implement these processes,
we have introduced 
a new parameter $R_{\rm grow}$ that defines the  'sphere of influence'
within which the dust accretion and destruction rates are calculated
for each dust particle based on the physical properties of 
gas around the dust particle.  Since the present results can possibly depend
strongly on this $R_{\rm grow}$, we need to investigate the possible
dependence.

Fig. B1 clearly shows that galactic SFHs
are qualitatively very similar  between the MW-type disk models
with four different $R_{\rm grow}$, though the SFRs at a given time step
are slightly different between the four models.
This is again encouraging, because we can estimate 
a reasonable $R_{\rm grow}$ from
the adopted gravitational softening length (corresponding
to the initial spatial resolution) of a simulation.
However,
this rather weak dependence on $R_{\rm grow}$ might be true only
for the present galaxy evolution simulations for isolated disks. 
Thus, we need to confirm whether this is the case for other models such as
galaxy formation models based on a $\Lambda$CDM cosmology.

\section{A brief summary of the new chemodynamical model}

\begin{figure*}
\psfig{file=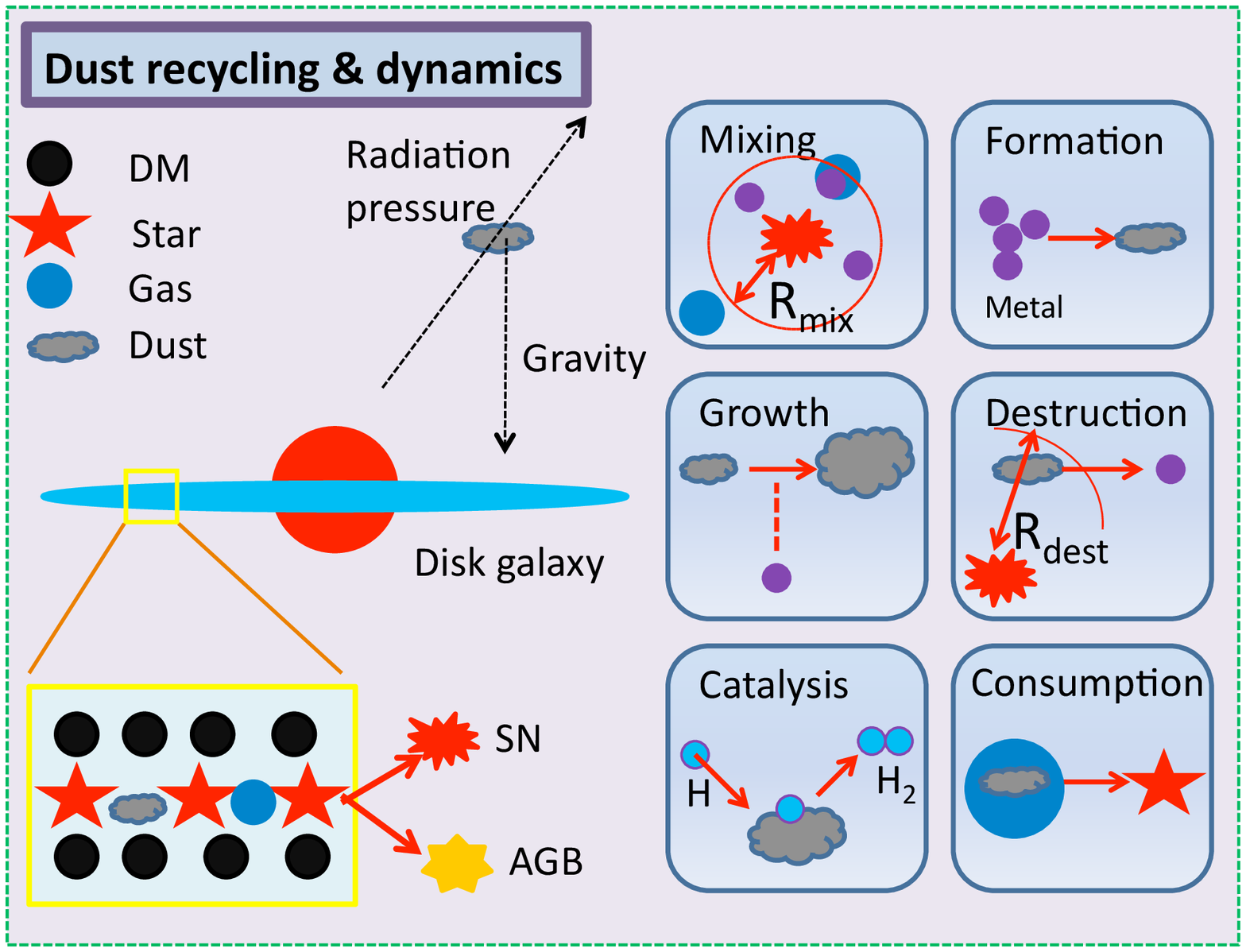,width=16.0cm}
\caption{
A cartoon representation of the new four-component chemodynamical model with
live dust particles. A galaxy is composed of dark matter (DM), stars, gas,
and dust, and the dust particles can move separately from gas in the new
model. The six key dust-related physical processes, i.e., dust (metal) mixing,
formation, growth, destruction, catalysis, and consumption are self-consistently
included in the Nbody+hydrodynamical simulations of galaxies. 
}
\label{Figure. C1}
\end{figure*}

Fig. C1 briefly summarizes the key elements of the new four-component chemodynamical
model adopted in the present numerical study of galaxy evolution. The six key physical
processes included in the code are (i) mixing of metals and dust ejected from SNe
and AGB stars, (ii) dust formation from condensation of metals in stellar winds of
SNe and AGB stars, (iii) dust growth through accretion of gas-phase metals onto
already existing dust grains, (iv) dust destruction by SNe, (v) ${\rm H_2}$ formation
on dust grains from neutral hydrogen ('catalysis'), and (vi) consumption of gas and
dust by star formation in molecular clouds in galaxies. 
The new model is quite different from previous standard three-component
chemodynamical model in the sense that (i) dust-related physical processes
are explicitly and self-consistently included and (ii) influences of dust evolution
on chemical evolution of gas-phase metals is considered.
The improved predictability of the new code is discussed extensively in the main
text.

\end{document}